\documentclass[reprint,superscriptaddress,nofootinbib,aps,prx]{revtex4-1}

\usepackage{amsfonts}
\usepackage{amsmath}
\usepackage{amssymb}
\usepackage{amsthm}
\usepackage{amsthm}

\newtheorem*{theorem}{Theorem}

\usepackage{bm}
\usepackage[hidelinks]{hyperref}
\usepackage{enumitem}

\usepackage{dcolumn}
\usepackage{epsfig}
\usepackage[outdir=./]{epstopdf} 
\usepackage{graphicx}
\usepackage{graphics}
\usepackage[latin1]{inputenc}
\usepackage{latexsym}
\usepackage{rotating}

\usepackage{xargs}
\usepackage[colorinlistoftodos,prependcaption,textsize=tiny]{todonotes}
\newcommandx{\improvement}[2][1=]{\todo[linecolor=Plum,backgroundcolor=Plum!25,bordercolor=Plum,#1]{#2}}
\newcommandx{\thiswillnotshow}[2][1=]{\todo[disable,#1]{#2}}

\usepackage{accents}

\newcommand{\ie}{\begin{equation}\begin{aligned}}
\newcommand{\fe}{\end{aligned}\end{equation}}
\newtheorem{thm}{Theorem}

\begin{document}
\hfill MIT-CTP/5508

\title{Obstructions to Gapped Phases from Non-Invertible Symmetries}

\author{Anuj Apte}
\affiliation{Kadanoff Center for Theoretical Physics \& Enrico Fermi Institute, Department of Physics, University of Chicago, Chicago, IL 60637}
\author{Clay C\'{o}rdova}
\affiliation{Kadanoff Center for Theoretical Physics \& Enrico Fermi Institute, Department of Physics, University of Chicago, Chicago, IL 60637}
\author{Ho Tat Lam}
\affiliation{Center for Theoretical Physics, Department of Physics,  Massachusetts Institute of Technology, Cambridge, MA 02139}

\begin{abstract}
Quantum systems in 3+1-dimensions that are invariant under gauging a one-form symmetry enjoy novel non-invertible duality symmetries encoded by topological defects.  These symmetries are renormalization group invariants which constrain dynamics.  We show that such non-invertible symmetries often forbid a symmetry-preserving vacuum state with a gapped spectrum.  In particular, we prove that a self-dual theory with $\mathbb{Z}_{N}^{(1)}$ one-form symmetry is gapless or spontaneously breaks the self-duality symmetry unless $N=k^{2}\ell$ where $-1$ is a quadratic residue modulo $\ell$. We also extend these results to non-invertible symmetries arising from invariance under more general gauging operations including e.g.\ triality symmetries.  Along the way, we discover how duality defects in symmetry protected topological phases have a hidden time-reversal symmetry that organizes their basic properties. These non-invertible symmetries are realized in lattice gauge theories, which serve to illustrate our results.  \end{abstract}

{
\let\clearpage\relax
\maketitle
}

%%%  I don't know why this works but it does
\makeatletter
\def\l@subsection#1#2{}
\def\l@subsubsection#1#2{}
\makeatother
\tableofcontents 

\section{Introduction}

The study of renormalization group flows and phase transitions has been a central topic in quantum field theory for the past half century.  A core conceptual idea is the organization of phases of field theories by their global symmetries and the realization of these symmetries on the ground state. This is the Landau paradigm for states of matter.  In this paper, we explore this framework for novel non-invertible duality symmetries.  

In its modern incarnation, the idea of symmetry has become intrinsically linked with topology. Each global symmetry of a theory may be understood as a co-dimension one defect operator (or as a domain wall in a spontaneously broken phase).  These symmetry defects are topological: continuous deformations of their positions which do not cross other operators leave their correlation functions invariant.

Over the last decade this idea has radically broadened in its scope and applicability, encompassing higher-form symmetry corresponding to invertible topological operators of general dimension \cite{Gaiotto:2014kfa}, higher-group symmetries which intertwine invertible topological operators of different dimensions \cite{Kapustin:2013uxa, Tachikawa:2017gyf, Cordova:2018cvg,Benini:2018reh}, and finally non-invertible symmetries  \cite{Bhardwaj2018,Tachikawa:2017gyf, Chang:2018iay, Kaidi:2021xfk, Choi:2021kmx}: the algebraic frontier where the defect operators are characterized by general higher fusion categories \cite{Gaiotto:2019xmp, Roumpedakis:2022aik, Bhardwaj:2022yxj, Choi:2022zal, Freed:2022qnc, Kaidi:2022cpf, Freed:2022iao}.  

In this work, we focus on non-invertible symmetries of (3+1)d theories, with the basic goal of determining when these symmetries are compatible with a unique vacuum state and a gapped spectrum.  As we describe below, in general we will find that such realizations of duality symmetries are impossible.  These obstructions are similar in spirit to the Lieb-Shultz-Mattis theorem \cite{Lieb:1961fr} which implies that certain (1+1)d lattice models are gapless or have degenerate ground states given the existence of certain types of symmetry.  Our results extend these ideas to the arena of higher-dimensional field theories invariant under a novel class of symmetries.

\subsection{Symmetry in Duality Invariant Theories}\label{sec:symintro}

 Examples of (3+1)d systems invariant under non-invertible symmetries may be constructed by starting from the class of quantum field theories which have a $\mathbb{Z}_{N}^{(1)}$ one-form global symmetry.  Given any such theory $\mathcal{Q}$, one can construct others by the following operations:
\begin{itemize}
\item $S$: Gauging the $\mathbb{Z}_{N}^{(1)}$ symmetry by coupling to a dynamical $\mathbb{Z}_{N}$ two-form gauge field $b$.  The resulting theory $S\mathcal{Q}$, then has an emergent dual $\mathbb{Z}_{N}^{(1)}$ global symmetry arising from the Wilson surface operators $\exp\left(\frac{2\pi i}{N}\oint b\right)$.
\item $T$: Stacking the theory $\mathcal{Q}$ with a minimal invertible theory for the $\mathbb{Z}_{N}^{(1)}$ global symmetry. In terms of a $\mathbb{Z}_{N}^{(1)}$ background gauge field $B$ this is expressed as 
\ie
\begin{cases}
\exp\left[\frac{2\pi i}{2N}\int_X \mathcal{P}(B)\right]\quad &N\text{ even ,}
\\
\exp\left[\frac{2\pi i(N+1)}{2N}\int_X \mathcal{P}(B)\right]\quad &N\text{ odd ,}
\end{cases}
\fe
where $\mathcal{P}(B)$ is a suitable quadratic function (see equation \eqref{sptdef} below).  
\end{itemize}
As an action on the set of all field theories with $\mathbb{Z}_{N}^{(1)}$ symmetry, $S$, and $T$ realize a discrete analog of the modular group and obey the equations:
\begin{equation}\label{modulargrp}
    S^{2}=C~, \quad (ST)^{3}=Y~,
\end{equation}
where $C$ is charge conjugation acting as $B\leftrightarrow -B$, and $Y$ represents stacking the original theory $\mathcal{Q}$ with an invertible field theory depending only on the spacetime manifold \cite{Gaiotto:2014kfa,Bhardwaj:2020ymp}.

\begin{figure}[ht!]
    \includegraphics[width = 0.45\textwidth]{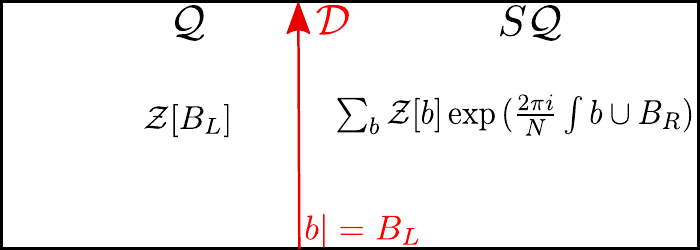}
    \caption{The definition of the duality defect $\mathcal{D}$ via gauging in half of spacetime.  The left region couples to a background two-form gauge field $B_L$ associated to the $\mathbb{Z}_{N}^{(1)}$ global symmetry.  In the right region this symmetry is gauged with dynamical field $b$. The right region recovers the $\mathbb{Z}_{N}^{(1)}$ global symmetry through the Wilson surface operators of $b$ which couple to the background field $B_R$. }
    \label{fig:dualitydef}
\end{figure}

By performing such a gauging procedure in half of space-time with Dirichlet boundary conditions separating the two halves, one obtains an interface separating two theories $\mathcal{Q}$ and $S\mathcal{Q}$ (see Figure \ref{fig:dualitydef}). As observed in \cite{Kaidi:2021xfk, Choi:2021kmx} when the theory is self-dual under gauging i.e.\ $\mathcal{Q}\cong SQ$, then this interface is a topological symmetry operator in $\mathcal{Q}$, a duality defect. Such duality defects generalize Kramers-Wannier duality lines from (1+1)d QFTs to this higher-dimensional setting.  More generally, one may also consider theories invariant under gauging operations built from composites of $S$ and $T$. For instance, invariance  $T^{-1}S$ leads to a triality defect \cite{Choi:2022zal}. 
  
These defects generalize the ordinary concept of symmetry.  In particular, their fusion is not defined by a group.  For instance, the duality defect $\mathcal{D}$ is invertible up to a condensate of one-form symmetry defects. Specifically, wrapping the defect $\mathcal{D}$ on a three-manifold $M$ and colliding it with its CPT conjugate $\overline{\mathcal{D}}$ one finds:
\begin{equation}\label{fusionrule}
    \mathcal{D}(M)\times \overline{\mathcal{D}}(M)\cong \frac{1}{N}  \sum_{S\in H_{2}(M,\mathbb{Z}_{N})} \hspace{-.1in}\exp\left(\frac{2\pi i}{N}\oint_{S} b\right)~,
\end{equation}
For a complete list of the fusion rules for the duality and triality defects see \cite{Choi:2022zal}.

A wide variety of non-trivial examples realizing these symmetries or related non-invertible defects have recently been constructed in the literature.  These include \cite{Koide:2021zxj,Choi:2021kmx,Kaidi:2021xfk,Cordova:2022rer,Benini:2022hzx,Roumpedakis:2022aik,Bhardwaj:2022yxj,Arias-Tamargo:2022nlf,Hayashi:2022fkw,Choi:2022zal,Kaidi:2022uux,Choi:2022jqy,Cordova:2022ieu,Antinucci:2022eat,Bashmakov:2022jtl,Damia:2022rxw,Damia:2022bcd,Moradi:2022lqp,Choi:2022rfe,Bhardwaj:2022lsg,Bartsch:2022mpm,Lin:2022xod,GarciaEtxebarria:2022vzq,Apruzzi:2022rei,Heckman:2022muc,Freed:2022qnc,Niro:2022ctq,Kaidi:2022cpf,Mekareeya:2022spm,Antinucci:2022vyk,Chen:2022cyw,Bashmakov:2022uek,Karasik:2022kkq,Cordova:2022fhg,Decoppet:2022dnz,GarciaEtxebarria:2022jky, Choi:2022fgx, Yokokura:2022alv, Bhardwaj:2022kot, Bhardwaj:2022maz, Bartsch:2022ytj, Hsin:2022heo, Das:2022fho, Heckman:2022xgu, Antinucci:2022cdi, Cordova:2022qtz}, which build on extensive previous investigations of non-invertible symmetries in $(1+1)d$ theories \cite{Verlinde:1988sn,Petkova:2000ip,Fuchs:2002cm,Frohlich:2004ef,Frohlich:2006ch,Feiguin:2006ydp,Frohlich:2009gb,Carqueville:2012dk,Aasen:2016dop,Bhardwaj:2017xup,Tachikawa:2017gyf,Chang:2018iay,Ji:2019ugf,Lin:2019hks,Thorngren:2019iar,Gaiotto:2020iye,Komargodski:2020mxz,Aasen:2020jwb,Chang:2020imq,Nguyen:2021naa,Thorngren:2021yso,Sharpe:2021srf,Huang:2021zvu,Huang:2021nvb,Vanhove:2021zop,Burbano:2021loy,Inamura:2022lun,Chang:2022hud,Lin:2022dhv,Robbins:2022wlr}.

\subsection{Phases of Duality Invariant Field Theories}

Like all notions of symmetry in quantum field theory, the duality defects described above can be used to constrain dynamics.
Below our main focus is on characterizing the interplay between duality invariant phases and the mass gap. 

 In general, given any symmetry one may always realize it in a spontaneously broken phase.  In the case of a discrete symmetry, such as the duality defects discussed here, a spontaneously broken phase means that there are multiple local vacua which are characterized by the presence of multiple topological local operators.\footnote{Here and below, by a local vacuum state we mean a ground state on $\mathbb{R}^{3}$ or equivalently $S^{3}$.}  The local physics in a given ground state is unconstrained, but the various ground states are related by the gauging operation.  This possibility can occur at a first order phase transition, where the duality defect forms the domain wall between different ground states.
 We discuss spontaneous symmetry breaking further in appendix \ref{app:SSB}, and highlight its occurrence in $\mathbb{Z}_{N}$ lattice gauge theory for small $N$ below.

In contrast with the scenario outlined above, symmetry preserving phases are tightly constrained.  In this situation there is a unique local topological operator (the identity).  Scale invariant phases arising at the end of renormalization group flows can be classified as follows:
\begin{itemize}[leftmargin=*]
    \item  Invertible Phases: The theory is gapped and invertible. Its partition function is a phase and there is a unique ground state on any spatial topology.   
    \item Topologically Ordered Phases: The theory is gapped, but described by a non-trivial topological quantum field theory (TQFT).  The ground state degeneracy depends on the spatial manifold.
    \item Gapless Phases: The mass gap vanishes and the theory is a free or interacting conformal field theory, with non-trivial power law correlation functions.
\end{itemize}

In the case of invertible symmetry (characterized by groups), the obstruction to the existence of an invertible realization is the 't Hooft anomaly.  However, anomalies can also obstruct the existence of topologically ordered phase. This phenomenon is referred to as \emph{symmetry enforced gaplessness} explored in condensed matter systems in \cite{Wang:2014lca, Wang:2016gqj, Sodemann:2016mib, Wang:2017txt, Kobayashi:2018yuk}, and through the lens of quantum field theory in \cite{Wang:2017loc, Wan:2018djl, Guo:2018vij, Kobayashi:2019lep, Cordova:2019bsd, Cordova:2019jqi, Thorngren:2020aph}.  

We generalize these considerations to (3+1)d duality invariant field theories by providing the first examples of TQFTs which are duality invariant without spontaneous symmetry breaking, and classifying all possible $N$ for which this can occur. This extends parallel results in (1+1)d \cite{Chang:2018iay} as well as the work of \cite{Choi:2021kmx, Choi:2022zal}, where duality invariant (3+1)d one-form symmetry protected topological (SPT) phases were classified. For the case of theories invariant under $S$ with the duality defect $\mathcal{D}$, our findings are enumerated in Table \ref{ns}.  

\begin{table}[]
\centering
\resizebox{0.48\textwidth}{!}{
\begin{tabular}{|c|l|}
\hline
SPT  & 2, 5, 10, 13, 17, 25, 26, 29, 34, 37, 41, 50, 53, 58, ...  \\ \hline
TQFT  & 4, 8, 9, 16, 18, 20, 25, 32, 36, 40, 45, 49, 50, 52, ...  \\ \hline
\end{tabular}
}
\caption{Possible $N \leq 60$ for $\mathbb{Z}_{N}^{(1)}$ symmetric gapped phases invariant under gauging ($S$). The top line enumerates those $N$ where an SPT exists.  These are integers where $-1$ is a quadratic residue modulo $N$ \cite{Choi:2021kmx}. The second line enumerates those $N$ where a duality invariant non-trivial TQFT exits.  In this case $N=k^{2}\ell$, where $-1$ is a quadratic residue modulo $\ell$. 
 A  duality symmetry preserving theory with $N$ that does not appear above is gapless. These include $N=3,6,7,11,12,...$.}
\label{ns}
\end{table}

Our basic method is to generalize the analysis of \cite{Cordova:2019bsd, Cordova:2019jqi} to the setting of non-invertible symmetries. The key idea is to examine the $\mathbb{Z}_{N}^{(1)}$ one-form symmetry action on line operators.  A distinguished role is played by the subgroup of $\mathbb{Z}_{N}^{(1)}$ that does not act faithfully, i.e.\ the subgroup of the symmetry with no associated charged objects.\footnote{Note that a one form symmetry which acts unfaithfully by linking with line operators may still have non-trivial correlation functions for instance via triple linking or other more intricate configurations \cite{Putrov:2016qdo, Hsin:2019fhf}.}  In a gapped phase, these operators admit topological boundary conditions.  If we further assume that there is no spontaneous symmetry breaking of the duality invariance, the effect of these unfaithfully acting surface operators can be reduced to insertions of the identity operator.  Enforcing duality invariance then leads to constraints on $N$ as elucidated in theorem \ref{thm:TQFT}.  We also provide a converse to our constraints on $N$ by explicitly constructing duality invariant TQFTs for all allowed values of $N$, generalizing the results of \cite{Thorngren:2020aph} to the setting of non-invertible symmetries.

Beyond simply reproducing the analysis of \cite{Choi:2021kmx} for SPT phases, our analysis of duality defects also reveals their physical properties.  In particular, we show that for a duality invariant SPT phase the world-volume theory of the defect $\mathcal{D}$  is a minimal Abelian TQFT \cite{Hsin2019} with a $\mathsf{T}$ invariant spectrum of anyons as analyzed in \cite{Delmastro2019}. This is analogous to the analysis presented in \cite{Kapustin:2014dxa, Hason2020, Yan2020}, where $\mathbb{Z}_{2}$ symmetry defects were shown to admit $\mathsf{T}$ symmetry with a map relating the bulk and defect anomalies.

Finally, we illustrate our results by considering $\mathbb{Z}_{N}$ lattice gauge theories. In these theories, topological phase transitions for $\mathbb{Z}_2,\mathbb{Z}_3$ and $\mathbb{Z}_4$ occur by spontaneous breaking of non-invertible self-duality, while for larger $N$ the duality invariant point is gapless. Thus, these first order phase transitions for small $N$ can be viewed as part of a generalized Landau paradigm by incorporating non-invertible symmetries. Several appendices summarize more technical material about quadratic Gauss sums, partition functions of topological two-form gauge theories, details of our argument classifying duality invariant TQFTs, and comments about models that spontaneously break duality symmetry.

\section{Invertible Phases}

In this section, we classify invertible field theories with duality symmetry which are invariant under combinations of the gauging operations $S$ and $T$ following \cite{Choi:2021kmx, Choi:2022zal}.  We also highlight how self-duality leads to an anti-unitary time-reversal symmetry acting on the duality defect $\mathcal{D}$, and comment on other physical aspects of the resulting symmetry defects. 

For simplicity, throughout this paper we will work on a smooth simply-connected Euclidean spacetime four-manifold $X$.  The manifold $X$ has the bi-linear intersection pairing in the middle dimension
\begin{equation}
    I: H^{2}(X;\mathbb{Z}) \times H^{2}(X;\mathbb{Z}) \to \mathbb{Z}~,
\end{equation}
defined by the cup product.  Associated to $I$ we introduce a quadratic function $\mathcal{P}(B)$ where $B\in H^{2}(X;\mathbb{Z}_{N})$ is a background field for the one-form symmetry and $\mathcal{P}$ depends on the parity of $N$ as:
\begin{equation}
    \mathcal{P}(B)=\begin{cases} \text{Pontryagin square}\in H^4(X,\mathbb{Z}_{2N})~  & N ~\text{even}~, \\
    B\cup B \in H^4(X,\mathbb{Z}_{N})~ & N~\text{odd}~.
    \end{cases}
\end{equation}

\subsection{Self-Dual SPT Phases}

The most general bosonic SPT phase for a $\mathbb{Z}_{N}^{(1)}$ one-form symmetry can be expressed as \cite{Kapustin2013}:
\ie\label{sptdef}
Z[B]=\exp\left[\int_X\frac{2\pi ip}{2N} \mathcal{P}(B)\right]~.
\fe
Here the integer $p$ characterizes the phase and is identified as $p\sim p+2N~$.  For $N$ even there are thus $2N$ distinct phases, while for $N$ odd, $p$ is further constrained to be even leading to $N$ distinct phases. 

The phase \eqref{sptdef} enjoys duality symmetry if it is invariant under the gauging operation $S$.  This can be directly checked by computing the partition function after gauging:\footnote{In \eqref{eq:gauged_partition_function_duality} one may also consider a more general bi-character by modifying the the weight of the $b\cup B$ term by an integer $\ell$ which is co-prime to $N$.  This does not modify the conclusions below.}
\ie\label{eq:gauged_partition_function_duality}
SZ[B]=\lambda_{N}^{-1}\sum_{b\in H^2(X,\mathbb{Z}_N)}\hspace{-.1in}Z[b]\exp\left[\int_{X}\frac{2\pi i}{N}b\cup B\right]~,
\fe
where $\lambda_{N}=\sqrt{|H^2(X,\mathbb{Z}_N)|}$. The right-hand side above is in general a non-invertible TQFT, and hence can only be equivalent to \eqref{sptdef} in the special case where all operators are trivialized by the equations of motion for $b$, given by
\begin{equation}\label{psolve}
    pb+B=0 ~\mod N~.
\end{equation}
If $p$ is co-prime to $N$, the above equation may be solved for $b$ restricting it to be a non-fluctuating classical background $B$.  Thus only in this case can gauging the SPT \eqref{sptdef} result in an invertible theory.  From now on we assume this condition. Substituting \eqref{psolve} back into the action \eqref{eq:gauged_partition_function_duality} leads to\footnote{ When $N$ is odd, we pick an even $(p)^{-1}_{N}$ so that the action is well-defined. If $(p)^{-1}_N$ is odd, we can redefine it by a shift of $N$ to obtain an even $(p)^{-1}_N$.}
\begin{equation}\label{eq:gauge_SPT_maintext}
    {SZ}[B]=G(I,p,N)\exp\left[-\int_X\frac{2\pi i\left(p\right)^{-1}_{\gamma(N)N}}{2N}\mathcal{P}\left({B}\right)\right]~.
\end{equation}
In the above, $G(I,p,N)$ is an overall $B$ independent phase defined by a Gauss sum for the intersection form ${I}$ (see appendix \ref{secgauss}):
\begin{equation}
    G(I,p,N)=\frac{1}{N^{\textrm{rk}(I)/2}}\sum_{x=0}^{N-1} \exp{\bigg(\frac{2 \pi i p}{2N} x^{T} I x\bigg)}~,
\end{equation}
where $\textrm{rk}(I)$ is the rank of $I$.

Let us focus first on the $B$ dependence of \eqref{eq:gauge_SPT_maintext}. We see that the gauged action has a quadratic dependence on $B$ with a map on $p$:
\begin{equation}
    p\mapsto -(p)^{-1}_{\gamma(N)N}~,  \hspace{.1in}\gamma(N) \equiv \begin{cases}
    1~\quad & \text{odd }N
    \\
    2~\quad & \text{even }N
\end{cases},
\end{equation}
where the notation $(\alpha)_\beta^{-1}$ denotes the inverse of $\alpha$ in the ring $\mathbb{Z}_\beta$ for co-prime integers $\alpha$ and $\beta$.  Invariance under the $S$ transformation therefore means that in the discrete group classifying the SPT, this map is the identity.  

Thus, a duality invariant SPT can exist only if there are solutions to the modular quadratic equation:
\begin{equation}\label{quadres1}
    p^{2}=-1~ \mod \gamma(N)N~.
\end{equation}
Moreover, when solutions to the above exist it is straightforward to check that the overall phase defined by the Gauss sum in \eqref{eq:gauge_SPT_maintext} is unity (see \eqref{gausstrivial}.)  As is well known, the existence of solutions to \eqref{quadres1} depends on the prime factorization of $N$. Thus, we reproduce the result of \cite{Choi:2021kmx}:
\begin{thm}\label{theorem:first}
    There exist invertible bosonic phases realizing the duality defect $\mathcal{D},$ i.e.\ invariant under the gauging operation $S$ if and only if $N$ is odd and a product of Pythagorean primes:
\begin{equation}
    N=p_{1}^{a_{1}}p_{2}^{a_{2}}\ldots p_{n}^{a_{n}}~, \quad \forall i~~~ p_{i}=1\text{ mod }4~.
\end{equation}
\end{thm}

For fermionic or equivalently spin theories, we can further refine the result above. When $N$ is even, using the fact that,
\begin{equation}
    \mathcal{P}(B)= B\cup w_{2}(X)~\mod 2~, 
\end{equation}
we deduce that on a spin manifold, $\mathcal{P}(B)$ is also even and hence there is an identification $p\sim p+N$. Then, the same steps as above imply that we have to solve the quadratic residue equation modulo $N$ for both odd and even $N$
\begin{equation}\label{spincriteria}
    p^{2}=-1~ \mod N~.
\end{equation}
Moreover in this case the overall phase defined by the Gauss sum is trivial due to Rokhlin's theorem (see below \eqref{bigform}). Therefore we again reproduce the result of \cite{Choi:2021kmx}:
\begin{thm}\label{theorem:second}
    There exist invertible spin phases invariant under the gauging operation $S$ if and only if $N$ a product of Pythagorean primes up to a single factor of two.
\end{thm}

\subsection{Physics of the Duality Defect}

It is instructive to analyze the physics of the duality defect $\mathcal{D}$. We focus below on the bosonic case, though analogous considerations hold for duality defects of fermionic SPTs.

\begin{figure}[h!]
    \includegraphics[width = 0.35\textwidth]{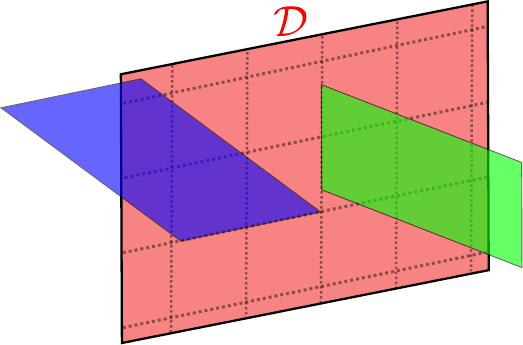}
    \caption{$\mathbb{Z}_{N}^{(1)} \times \mathbb{Z}_{N}^{(1)}$ symmetry of $\mathcal{D}$ arises from ending of one-form symmetry defects from $\mathcal{Q}$ (purple) and  $S\mathcal{Q}$ (green).}
    \label{enhanced}
\end{figure}

With general bulk dynamics, the symmetry defect has a $\mathbb{Z}_{N}^{(1)}\times \mathbb{Z}_{N}^{(1)}$ symmetry which arises from bulk one-form symmetry defects ending on the duality defect (see Figure \ref{enhanced}). This ending is possible because the duality defect is defined by Dirichlet boundary conditions for the dynamical gauge field (see Figure \ref{fig:dualitydef}). 

Physically, the one-form symmetries of $\mathcal{D}$ define Abelian anyons within $\mathcal{D}$. The self-braiding of each $\mathbb{Z}_{N}^{(1)}$ factor is theory dependent, but the braiding between the generators of each $\mathbb{Z}_{N}^{(1)}$ factor is fixed by the pairing of $b$ and $B$ in the gauging defining the defect (See Figure \ref{fig:dualitydef}).\footnote{This generalizes an analogous statement in (1+1)d theories with Tambara-Yamagami fusion category symmetry where the bi-character encodes the anomaly between left and right symmetries of the defect \cite{Chang:2018iay, Wang2019,Thorngren:2021yso}.}

\begin{figure}[h!]
    \includegraphics[width = 0.45\textwidth]{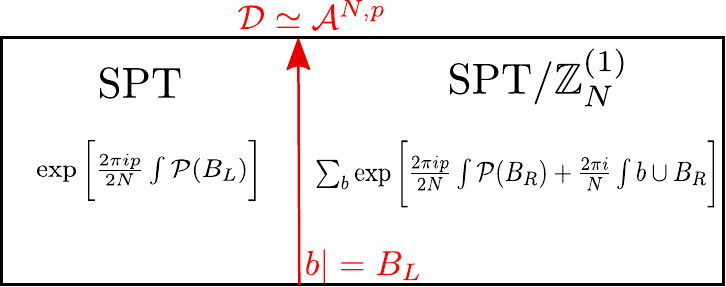}
    \caption{The duality defect $\mathcal{D}$ in a bulk SPT phase. The defect $\mathcal{D}$ is a well-defined (2+1)d TQFT which is indentified with a minimal abelian TQFT.}
    \label{fig:dualitydefspt}
\end{figure}

When the bulk dynamics are invertible the theory on the duality defect is a well-defined (2+1)d TQFT and the bulk SPT phase on the left and the right specify the anomaly of this system. Moreover the construction of the defect by gauging in a half space identifies the (2+1)d defect theory as a minimal Abelian TQFT $\mathcal{A}^{N,p}$ \cite{Hsin2019} (see Figure \ref{fig:dualitydefspt}). Such a theory has a spectrum of lines which are exactly given by the abelian anyons defining the one-form symmetry.  

The theory $\mathcal{A}^{N,p}$ has a single $\mathbb{Z}_{N}^{(1)}$ symmetry generated by an abelian anyon $a$.  Thus, the left and right bulk symmetry surfaces must each end on these abelian anyons but with a possible difference in the choice of generating line.  To deduce the relationship between left and right, we consider the equation of motion \eqref{psolve} and interpret $B$ and $b$ as the fields on the wall sourcing the right and left symmetries.  When \eqref{psolve} is not solved the partition function of the total system including the defect vanishes. This means that the quantity appearing in the equations of motion sources charged abelian anyons wrapping non-trivial cycles in the defect.  Correspondingly, when the equation of motions hold no charged lines are inserted. Hence, the left and right surfaces end on the lines: 
\begin{equation}\label{leftrightmap}
\text{Left}:~ a \hspace{.2in} \text{Right}:~ a^{(p)^{-1}_N}~. 
\end{equation}

This relation between the left and the right symmetries \eqref{leftrightmap} may be further understood by observing that in an invertible phase, the duality defect $\mathcal{D}$ has an antiunitary time-reversal symmetry, $\mathsf{T}$. Note that this is true even though for general $p$ the bulk SPT defined in \eqref{sptdef} does not have $\mathsf{T}$ symmetry.

\begin{figure*}[tb!]
\centering
    \includegraphics[width = 0.8\textwidth]{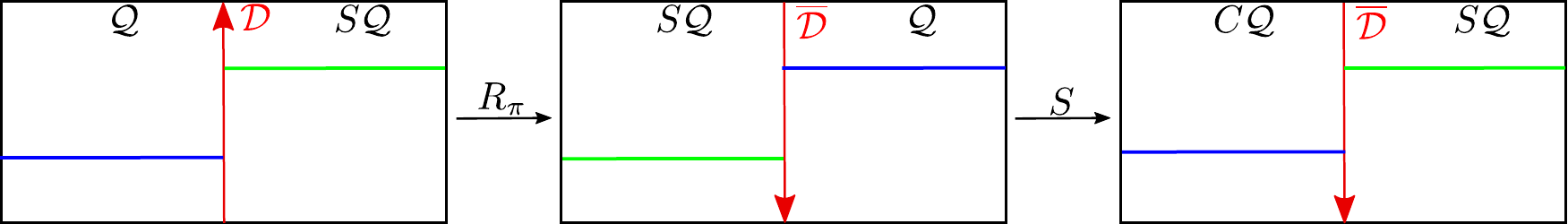}
    \caption{$\mathsf{T}$ symmetry of the duality defect separating phases $\mathcal{Q}$ and $S\mathcal{Q}$. $R_{\pi}$ combined with gauging of one-form symmetry in all of spacetime maps the boundary conditions to themselves, but reverses the orientation of $\mathcal{D}$.  Hence this composite operation defines a time-reversal symmetry $\mathsf{T}$ of the duality defect worldvolume.}
    \label{doubling}
\end{figure*}

 To argue for this conclusion we proceed analogously to \cite{Kapustin:2014dxa, Hason2020, Yan2020}.  Consider the theories $\mathcal{Q}$ and $S\mathcal{Q}$ separated by the duality defect as in Figure \ref{doubling}. A rotation by $\pi$ along an axis in the defect, which we denote by $R_{\pi}$, swaps the theories $\mathcal{Q}$ and $S\mathcal{Q}$. By composing this operation with gauging of the one-form symmetry in all of space-time i.e.\ $S$ we obtain a symmetry $\mathsf{T}$
\begin{equation}
    \mathsf{T} \equiv S \cdot R_{\pi}~.
\end{equation}
By construction this leaves the boundary conditions invariant and therefore acts on the duality defect. Moreover since it reverses the defect orientation it is a time-reversal symmetry.\footnote{More generally when the bulk dynamics is not invertible, we expect that the duality defect $\mathcal{D}$ still enjoys an antiunitary non-invertible symmetry analogous to those recently discussed in \cite{Choi:2022rfe}.} 

\begin{figure}[h!]
    \includegraphics[width = 0.45\textwidth]{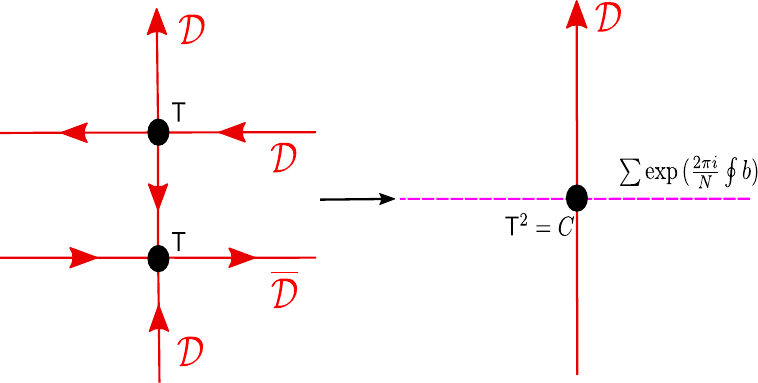}
 \caption{The $\mathsf{T}$ symmetry of the duality defect $\mathcal{D}$ is realized by intersecting duality defects (shown as black dots).  At each such intersection the duality defect orientations reverse.  Colliding two such junctions leads to the bulk fusion algebra of $\mathcal{D}$ with its orientation reversal $\overline{\mathcal{D}}$ resulting in a condensation of one-form symmetry surfaces shown in magenta.  Within the duality defect worldvolume, this restricts to a condensation of abelian anyons that produces the charge conjugation symmetry $C$.}\label{t2fig}
\end{figure}

We can also deduce the unitary symmetry  $\mathsf{T}^{2}$ from the fusion algebra of the duality defect \eqref{fusionrule}.  Within the wall the $\mathsf{T}$ symmetry defect is realized by intersecting $\mathcal{D}$ with itself.  Since this is a time-reversal symmetry, at each such intersection the orientations of the defects reverse (see Figure \ref{t2fig}).  If we now collide two such junctions we obtain the fusion of $\mathcal{D}\times \overline{\mathcal{D}}$ resulting in a condensation of one-form symmetry surfaces.  Such surfaces come in two varieties:
\begin{itemize}
    \item Surfaces parallel to the vertical duality defect $\mathcal{D}$ in Figure \ref{t2fig}.  These surfaces are simply absorbed by the duality defect.  Indeed, $\mathcal{D}$ is characterized by Dirichlet boundary conditions (see Figure \ref{fig:dualitydefspt}), and therefore absorbs one-form symmetry defects contained completely within it. 
    \item One-form symmetry surfaces that intersect the vertical duality defect $\mathcal{D}$ in Figure \ref{t2fig}.  These are non-trivial and give rise to abelian anyons within the duality defect.  Taking into account the identification of generators, we see that a generator for the bulk one-form symmetry yields an abelian anyon $a^{1+(p)^{-1}_N}$
    within $\mathcal{D}.$
\end{itemize}

From these observations, we deduce that the operator $\mathsf{T}^{2}$ in $\mathcal{D}$ is defined by a condensate of abelian anyons \cite{Roumpedakis:2022aik}.  For instance taking the surface supporting $\mathsf{T}$ to be a torus $M$ we find:
\begin{equation}\label{surfcond1}
    \mathsf{T}^{2}=\frac{1}{N}\sum_{\gamma \in H_{1}(M, \mathbb{Z}_{N})} a^{1+(p)^{-1}_N}(\gamma)~.
\end{equation}
Although condensation operators like the one appearing above are frequently non-invertible, in fact in this special case the condensation is invertible and is identified with the charge conjugation symmetry $C$ that maps $a^r$ to $a^{-r}$.

To demonstrate this we note that as a consequence of \eqref{quadres1}, when $N$ is odd, $1+(p)^{-1}_N$ is co-prime to $N$.\footnote{ Using \eqref{quadres1}, we have $(1+(p)^{-1}_N)(2)^{-1}_N(1-(p)^{-1}_N)=1$ mod $N$ for odd $N$ and therefore $1+(p)^{-1}_N$ is co-prime to $N$.}  Therefore, the condensation defect above is equivalent to a sum over all abelian anyons: 
\begin{equation}\label{surfcond2}
    \mathsf{T}^{2}=\frac{1}{N}\sum_{\gamma \in H_{1}(M, \mathbb{Z}_{N})} a(\gamma)~.
\end{equation}
Now we follow \cite{Roumpedakis:2022aik} and evaluate the action of $\mathsf{T}^{2}$ on a line $a^{s}$. Let $\alpha, \beta \in H_{1}(M,\mathbb{Z})$ denote a basis for the one-cycles on the torus with intersection form $\alpha \cap \beta =1$.  We can split a general line on a cycle $\gamma$ as:
\begin{equation}\label{splitgamma}
    a(x \alpha+y \beta)=\exp\left(\frac{2\pi i p x y}{2N}\right)a^{y}(\beta)a^{x}(\alpha)~,
\end{equation}
where $x,y\in \mathbb{Z}$, and above we have used that the spin of the line $a^{r}$ is  fixed as:
\begin{equation}\label{spinform}
h(a^r) = \frac{p r^2}{2N}~.
\end{equation}
We consider a geometry where the anyon $a^{s}$ wraps the $\beta$ cycle at the center of the solid torus whose boundary is the support $M$ of $\mathsf{T}^{2}.$ Then the lines wrapping $\alpha$ braid with the central anyon, while those wrapping $\beta$ fuse with it.  Whence using \eqref{splitgamma}:
\begin{eqnarray}
    \mathsf{T}^{2} \left( a^{s}\right) & =& \frac{1}{N}\sum_{x,y=0}^{N-1} \exp\left(\frac{2\pi i p x y}{2N}\right)a^{y}(\beta)a^{x}(\alpha) a^{s}(\beta)~, \nonumber\\
    & = & \frac{1}{N}\sum_{x,y=0}^{N-1} \exp\left(\frac{2\pi i p x y}{2N}+\frac{2\pi i p x s}{N}\right)\left(a^{y+s}\right)~.
\end{eqnarray}
The sum above over $x$ now enforces
\begin{equation}
    \frac{py}{2}+p s=0~, ~\mod N~.
\end{equation}
And since $N$ is odd and $p$ and $N$ are co-prime, we can solve for $y$ to find simply:
\begin{equation}
     \mathsf{T}^{2} \left( a^{s}\right)=a^{-s}~.
\end{equation}
This is exactly the expected action of charge conjugation on lines.  In summary we have derived the algebra:\footnote{For even $N$, charge conjugation is instead the condensation defect generated by $a^{2}$. This matches \eqref{surfcond1} because \eqref{spincriteria} implies that $(1+(p)^{-1}_N)(1-(p)^{-1}_N)=2$ mod $N$ and therefore $1+(p)^{-1}_N$ is even and $(1+(p)^{-1}_N)/2$ is co-prime to $N/2$.}
\begin{equation}
    \mathsf{T}^{2}=C~.
\end{equation}

The symmetry $\mathsf{T}$ that we have identified provides another explanation of several of the features of the duality defect.  As is evident from Figure \ref{doubling}, $\mathsf{T}$ exchanges the left and right bulk one-form symmetries.  Since these couple differently to the defect as in \eqref{leftrightmap}, we conclude that $\mathsf{T}$ acts on the anyons of the defect theory $\mathcal{A}^{N,p}$ as:
\begin{equation}
    \mathsf{T}(a)=  a^{(p)^{-1}_N} 
    ~.
\end{equation}
We can see that this is a consistent action of time-reversal by recalling that the spin of the anyon $a^{r}$ in \eqref{spinform}.  Indeed, $\mathsf{T}$ changes the sign of the spin precisely when $p$ obeys $p^{2}=-1 \mod N$.  This reproduces the condition \eqref{psolve} necessary for the existence of a duality invariant SPT.  However, now we see that it is a consequence of the $\mathsf{T}$ symmetry of the (2+1)d duality defect TQFT in agreement with analysis of \cite{Delmastro2019}.

Finally, let us comment on the chiral central charge $c$ of the duality defect.  This may be computed by summing over the spins of the distinct lines \eqref{spinform}.  Recalling that $pN$ is even (and hence $\mathcal{A}^{N,p}$ is bosonic), the distinct lines correspond to $a^{r}$ for $r=\{0,\ldots,N-1\}$.  This yields:
\begin{equation}\label{sumform}
    \exp{\bigg(\frac{2 \pi i c} {8}\bigg)} = \frac{1}{\sqrt{N}}\sum_{r=0}^{N-1} e^{\frac{2 \pi i p}{2 N} r^2} = G([+1],p,N)~,
\end{equation}
where the right-hand side is the quadratic Gauss sum for the intersection form of $\mathbb{CP}^2$.  Pragmatically, this is also the same phase factor that we encounter in \eqref{eq:gauge_SPT_maintext} when we integrate out the dynamical field $b$. 

As remarked above (and proven in \eqref{gausstrivial}), when the conditions for the existence of the duality defect are met, namely when $p^{2}=-1$ mod $N$ for odd $N$, the Gauss sum above is trivial and hence the chiral central charge vanishes modulo eight.  In particular, this implies that up to possibly stacking with a properly quantized bosonic gravitational Chern-Simons term, we deduce that the chiral central charge of the duality defect $\mathcal{A}^{N,p}$ vanishes.

In fact, the condition $c=0$ is also compatible with the time-reversal invariance of the duality defect, since $\mathsf{T}$ relates left and right moving chiral edge modes.  Thus we see how the $\mathsf{T}$ symmetry of the duality defect explains and unifies many of its physical properties.

\subsection{General Duality Invariant Invertible Phases}

It is straightforward to extend the previous analysis to invertible phases invariant under compositions of gauging ($S$) and stacking ($T$) operations defined in section \ref{sec:symintro}.  Let us now consider the composite operation $T^{-p'}S$. In particular, the case $p'=\pm 1$ or $p'=\pm (1+N)$, for even or odd $N$ respectively give rise to the existence of triality defects studied in \cite{Choi:2022zal}.

Acting on the general invertible theory defined in \eqref{sptdef}, the operation $T^{-p'}S$ leads to the action
\ie\label{eq:gauged_partition_function_general}
&(T^{-p'}SZ)[B] \\
&=\lambda_{N}^{-1}\sum_{b\in H^2(X,\mathbb{Z}_N)}\hspace{-.1in}Z[b]\exp\left[\int_{X}\frac{2\pi i}{N}b\cup B - \frac{2\pi ip'}{2N} \mathcal{P}(B)\right]~.
\fe
Using the equations of motion for $b$, we can follow the same steps that we did to derive \eqref{eq:gauge_SPT_maintext} to obtain
\ie\label{eq:general_partition_function_SPT}
&(T^{-p'}SZ)[B] \\
&=G(I,p,N)\exp\left[-\int_X\frac{2\pi i\left(p'+(p)_{\gamma(N)N}^{-1}\right)}{2N}\mathcal{P}\left({B}\right)\right]~,
\fe
If the theory $T^{-p'}S\mathcal{Q}$ is equivalent to the original theory $\mathcal{Q}$ (up to a gravitational counter-term), then the dependence on $B$ must match. Imposing this condition leads to the following theorem \cite{Choi:2022zal}:
\begin{thm}\label{thm:SPT_trialitybos}
There exists a bosonic SPT that is invariant under $T^{-p'}S$ gauging a $\mathbb{Z}_N^{(1)}$ one-form symmetry, if and only if there exists a solution $p$ to 
\ie \label{eq:triality_condition_bosonic} p(p+p')=-1 \mod \gamma(N)N~.
\fe 
\end{thm} 
Analogously, for spin theories we have \cite{Choi:2022zal}:
\begin{thm}\label{thm:SPT_trialityspin}
There exists a spin SPT that is invariant under $T^{-p'}S$ gauging a $\mathbb{Z}_N^{(1)}$ one-form symmetry, if and only if there exists a solution $p$ to 
\ie \label{eq:triality_condition_spin}
p(p+p')=-1\mod N~.
\fe 
\end{thm}

We now restrict our attention to triality invariant SPTs. For even $N$, this means that the SPT is invariant under $T^{-1}S$ gauging the $\mathbb{Z}_N^{(1)}$ one-form symmetry. Such a (spin) SPT does not exist because there is no solution to the conditions \eqref{eq:triality_condition_bosonic} or \eqref{eq:triality_condition_spin} when $p'=1$ and $N$ is even. For odd $N$, $p'=-(1+N)$ and such SPTs may exist.\footnote{Inspecting equation \eqref{eq:triality_condition_bosonic}, we deduce that solutions exist if and only if -3 is a quadratic residue modulo $N$ \cite{Choi:2022zal}.}

As in the case of SPTs invariant under the $S$ operation, it is instructive to analyze the physics on the triality defect defined by doing $T^{-(1+N)}S$ gauging in half of the spacetime.
The theory on the triality defect is the minimal Abelian TQFTs $\mathcal{A}^{N, p}\times \mathcal{A}^{N,N+1}$. The first minimal TQFT comes from $S$ gauging the SPT in half of the spacetime while the second one is introduced to cancel the anomaly inflow from the $T^{-(1+N)}$ operation on the right. We can evaluate the chiral central charge of the TQFT on the defect by summing over lines as in \eqref{sumform} yielding:
\ie \label{chiraltrial}
    \exp{\bigg(\frac{2 \pi i c} {8}\bigg)} = G([+1],p,N)G([+1],1+N,N)
    ~.
\fe
The first Gauss sum is the same phase factor that shows up in \eqref{eq:general_partition_function_SPT}.
The Gauss sum in \eqref{chiraltrial} does not equal to unity in general but it is independent of the integer $p$ characterizing the triality invariant SPT. Using \eqref{eq:triality_Gauss_sum}, we get
\ie
    \exp{\bigg(\frac{2 \pi i c} {8}\bigg)} = \begin{cases} 1  & N \equiv 1 \mod{4}~, \\
    -1  & N \equiv 3 \mod{4}~.
    \end{cases}
\fe
 Since the Gauss sum is always an eighth root of unity the chiral central charge of the symmetry defect may be cancelled by stacking with a well quantized spin$_{\mathbb{C}}$ gravitational Chern-Simons term.  

Therefore, for triality invariant SPT phases, the symmetry defects implementing the gauging operation have a vanishing chiral central charge modulo the addition of invertible (2+1)d gravitational theories.

\section{Topologically Ordered Phases}

In this section, we construct examples of (3+1)d duality symmetry preserving  TQFTs which are invariant under combinations of $S$ and $T$ gauging operations  of a $\mathbb{Z}_N^{(1)}$ one-form symmetry.  
Without loss of generality, we consider only TQFTs that have a unique local vacuum.
After these explicit constructions, we prove that for values of $N$ that are not realized by these explicit examples, such TQFTs do not exist.

\subsection{Examples}

As a starting example, consider the case when $N=k^2$ is a perfect square. Then, a $\mathbb{Z}_k$ gauge theory is invariant under gauging a $\mathbb{Z}_N^{(1)}$ one-form symmetry that couples to the background gauge field $B$ as
\ie
Z[B]&=\lambda_k^{-1}\sum_{b\in H^2(X,\mathbb{Z}_k)}\exp\left[\frac{2\pi i}{k}  \int_Xb\cup B\right]
\\
&=\lambda_k\, \delta (B\text{ mod }k)~,
\fe
where $\lambda_k=\sqrt{|H^2(X,\mathbb{Z}_k)|}$.
The $\mathbb{Z}_N^{(1)}$ one-form symmetry, in particular its $\mathbb{Z}_k^{(1)}$ subgroup, does not act faithfully on line operators. It is straightforward to check that gauging the $\mathbb{Z}_N^{(1)}$ one-form symmetry leads to the same partition function as the original $\mathbb{Z}_k$ gauge theory.
Using the half space gauging construction, we find that the duality defect factorizes into the product of Dirichlet boundary conditions for the $\mathbb{Z}_k$ two-form gauge fields on the two sides.

More generally, consider the case when $N=k^2\ell$ for some integers $k$ and $\ell$, and -1 is a quadratic residue of $\ell$. Then, there exists a (spin) TQFT that is invariant under gauging a $\mathbb{Z}_N^{(1)}$ one-form symmetry. 
Let $p$ be a solution to $p^2+1=0$ mod $\ell$. In particular, this implies that $p$ and $\ell$ are co-prime.

Let us first consider the case when $\ell$ is odd. Then solutions exist if and only if $\ell$ is a product of Pythagorean primes.  
The TQFT invariant under gauging is a $\mathbb{Z}_k$ gauge theory stacked with a $\mathbb{Z}_N^{(1)}$ one-form SPT described by the following partition function\footnote{For the expressions to be well-defined, we pick an even solution to $p^2+1=0$ mod $\ell$. If the solution $p$ is odd, we can get an even solution by redefining $p$ to $p+\ell$.}
\ie\label{eq:TQFT_ansatz}
&Z[B]
\\
&=\lambda_k^{-1}\sum_{b\in H^2(X,\mathbb{Z}_{k})}\exp\left[2\pi i\int_X\left(\frac{ p}{2N}\mathcal{P}(B)+\frac{1}{k}b\cup B\right)\right]
\\
&=\lambda_k\delta(B\text{ mod }{k})\exp\left[\frac{2\pi i p}{2\ell}\int_X\mathcal{P}\left(\frac{B}{k}\right)\right]~.
\fe
Gauging the $\mathbb{Z}_N^{(1)}$ one-form symmetry, we obtain a partition function identical to \eqref{eq:TQFT_ansatz}
\ie
&SZ
[B]
\\
&=\lambda_{k\ell}^{-1}\sum_{b\in H^2(X,\mathbb{Z}_{k\ell})}\exp\left[2\pi i\int_X\left(\frac{ p}{2\ell}\mathcal{P}(b)+\frac{1}{k\ell} b\cup B\right)\right]
\\
&=\lambda_k\delta(B\text{ mod }{k})G(I,p,\ell)\exp\left[-\frac{2\pi i(p)_\ell^{-1}}{2\ell}\int_X\mathcal{P}\left(\frac{B}{k}\right)\right]
\\
&=\lambda_k\delta(B\text{ mod }{k})\exp\left[\frac{2\pi ip}{2\ell}\int_X\mathcal{P}\left(\frac{B}{k}\right)\right]=Z[B]~.
\fe
In the second equality, we used \eqref{eq:gauge_SPT_maintext} for the odd $\ell$ case since $p$ and $\ell$ are co-prime, and we pick an even $(p)^{-1}_\ell$. In the last equality, we simplified the expression using the fact that $G(I,p,\ell)=1$ when $p^2+1=0$ mod $\ell$ and and
$(p)_\ell^{-1}/2=-p/2$ mod $\ell$. Note that the TQFT \eqref{eq:TQFT_ansatz}, as a bosonic TQFT, is already invariant under gauging.

Now consider the case when $\ell$ is even. Then solutions exist if $\ell/2$ is odd and $\ell/2$ is a product of Pythagorean primes. Again, let us consider the theory \eqref{eq:TQFT_ansatz}. Using \eqref{eq:gauge_SPT_maintext} for the even $\ell$ case, gauging leads to
\ie\label{eq:TQFT_gauging_even_ell}
SZ
[B]=&\,\lambda_{k}\delta(B\text{ mod }k)G(I,p,\ell)
\\
&\,\times \exp\left[-\frac{2\pi i(p)_{2\ell}^{-1}}{2\ell}\int_X\mathcal{P}\left(\frac{B}{k}\right)\right]~.
\fe
As a spin TQFT, the theory is exactly invariant under gauging the $\mathbb{Z}_N^{(1)}$ one-form symmetry since $(p)_{2\ell}^{-1}=-p\text{ mod }\ell$ and $G(I,p,\ell)=1$ on spin manifolds. However, as a bosonic TQFT, the theory cannot be invariant under gauging because there is no solution to $(p)_{2\ell}^{-1}=-p\text{ mod }2\ell$.

\subsection{Constraints on TQFTs}

So far, we have explicitly constructed (spin) TQFTs that have a unique local vacuum and are invariant under gauging a $\mathbb{Z}_N^{(1)}$ one-form symmetry for some particular $N$. We now prove that such TQFTs do not exist for the values of $N$ that have not been realized in the last subsection. This means that
\begin{thm}\label{thm:TQFT}
There exists a (spin) TQFT, that has a unique local vacuum and is invariant under gauging a $\mathbb{Z}_N^{(1)}$ one-form symmetry up to a gravitational counterterm $e^{i\Omega(X)}$, if and only if $N=k^2\ell$ for some integers $k$ and $\ell$ and $-1$ is a quadratic residue of $\ell$ (or equivalently $\ell$ is a product of Pythagorean primes up to a factor of two). If such a (spin) TQFT exists, $e^{i\Omega(X)}=1$ for every simply-connected smooth spin 4-manifold $X$.
\end{thm}
 
We now present a brief sketch of the proof, while the details are fleshed out in appendix \ref{app:TQFT}.

In general, the $\mathbb{Z}_N^{(1)}$ one-form symmetry can act unfaithfully on line operators \textit{i.e.}\ some symmetry surface operators link trivially with all line operators. Denote the unfaithful subgroup by $\mathbb{Z}_M^{(1)}$. Then, we have $N=Mk$ for some integers $M$ and $k$. If the theory is a TQFT, after gauging the $\mathbb{Z}_N^{(1)}$ one-form symmetry, the $\hat{\mathbb{Z}}_k^{(1)}$ subgroup of the dual $\hat{\mathbb{Z}}_N^{(1)}$ one-form symmetry necessarily act unfaithfully on line operators because their symmetry surface operators admit topological boundary conditions provided by the topological line operators charged under the original $\mathbb{Z}_N^{(1)}$ one-form symmetry. Using these topological boundary conditions, we can unlink the dual $\hat{\mathbb{Z}}_k^{(1)}$ symmetry operators and any line operator. Since the TQFT is invariant under gauging the $\mathbb{Z}_N^{(1)}$ one-form symmetry, the unfaithful subgroup of the dual $\hat{\mathbb{Z}}_N^{(1)}$ one-form symmetry should be $\hat{\mathbb{Z}}_M^{(1)}$, which should include the unfaithful $\hat{\mathbb{Z}}_k^{(1)}$ one-form symmetry as a subgroup. This implies that $M$ is divisible by $k$ and $N=k^2\ell$ for some integer $\ell$.

In a TQFT, symmetry operators of an unfaithful one-form symmetry admit topological boundary conditions \cite{Cordova:2019bsd}. In our case, this means that we can cut open the symmetry surface operators of the unfaithful $\mathbb{Z}_M^{(1)}$ one-form symmetry and shrink them to local operators. Consider placing the TQFT on $X=S^2\times S^2$ and turning on the background gauge field $B=(B_1,B_2)\in H^2(S^2\times S^2,\mathbb{Z}_M)=\mathbb{Z}_M^2$ for the unfaithful $\mathbb{Z}_M^{(1)}$ one-form symmetry. By shrinking the symmetry operators to local operators at their intersections, we argue that the partition function takes the form 
\ie\label{eq:partition_function_ZM}
Z[B_1,B_2]=Z[0,0]\exp\left(\frac{2\pi i p}{M}B_1 B_2\right)~.
\fe
The partition function $Z[0,0]$ on $S^2\times S^2$ is a positive number because of unitarity \cite{Cordova:2019bsd}.

Using the fact that the TQFT is invariant under gauging the $\mathbb{Z}_N^{(1)}$ one-form symmetry (upto a gravitational counterterm), we show that the partition function \eqref{eq:partition_function_ZM} obeys a constraint
\ie\label{eq:constraint_ZM}
&\frac{1}{{k\ell}}\sum_{b_{1,2}=1}^{k\ell} Z[b_1,b_2]\exp\left(\frac{2\pi i}{\ell}(b_1C_2+b_2 C_1)\right)
\\
&=\frac{1}{{k}}\sum_{b_{1,2}=1}^kZ[\ell b_1+C_1,\ell b_2+C_2]\times e^{i\Omega(S^2\times S^2)}~,
\fe
where $C=(C_1,C_2)\in H^2(S^2\times S^2,\mathbb{Z}_\ell)=\mathbb{Z}_\ell^2$ and $e^{i\Omega(S^2\times S^2)}$ is a phase independent of $(C_1,C_2)$. First line of the equality is the partition function of the TQFT after gauging the unfaithful $\mathbb{Z}_M^{(1)}$ one-form symmetry with the dual $\tilde{\mathbb{Z}}_\ell^{(1)}$ background gauge field $C$ turned on. The same partition function can also be obtained from the $\mathbb{Z}_N^{(1)}$-gauged theory by gauging the $\hat{\mathbb{Z}}_k^{(1)}$ subgroup of the dual $\hat{\mathbb{Z}}_N^{(1)}$ one-form symmetry with the background gauge field $C$ for the dual $\hat{\mathbb{Z}}_M^{(1)}/\hat{\mathbb{Z}}_k^{(1)}$ one-form symmetry turned on. Since the $\mathbb{Z}_N^{(1)}$-gauged TQFT share the same partition function as the original TQFT up to a gravitation counterterm $e^{i\Omega(X)}$,
\ie
Z[B]=(SZ)[B]e^{-i\Omega(X)}~,
\fe
we have the equality \eqref{eq:constraint_ZM}.

In the end, the partition function \eqref{eq:partition_function_ZM} together with the constraint \eqref{eq:constraint_ZM} implies that $-1$ is a quadratic residue of $\ell$. We also learned that $e^{i\Omega(S^2\times S^2)}=1$. Similar reasoning holds on every simply-connected smooth spin 4-manifold $X$. This leads to the same constraint on $\ell$ and gives $e^{i\Omega(X)}=1$. This concludes the proof for theorem \ref{thm:TQFT}.

\subsection{General Duality Invariant TQFTs}

It is straightforward to extend the results to more general non-invertible symmetries associated to the invariance of $T^{-p'}S$ gauging, including the triality symmetry.

Generalizing the argument used in the proof of theorem \ref{thm:TQFT}, in appendix \ref{app:TQFT} we prove the following theorem:
\begin{thm}\label{thm:TQFT_triality}
There exists a (spin) TQFT, that has a unique local vacuum and is invariant under $T^{-p'}S$ gauging a $\mathbb{Z}_N^{(1)}$ one-form symmetry up to a gravitational counterterm $e^{i\Omega(X)}$, if and only if $N=k^2\ell$ for some integers $k$ and $\ell$ and there exists a solution $p$ to \ie\label{eq:solution}
p(p+p')+1=0\text{ mod }\ell~.
\fe
If such a (spin) TQFT exists, $e^{i\Omega(X)}=1$ for every simply-connected smooth spin 4-manifold $X$.
\end{thm}
When $N$ obeys the condition, we can explicitly construct examples of spin TQFTs that are invariant under $T^{-p'}S$ gauging of the $\mathbb{Z}_N^{(1)}$ one-form symmetry.
This spin TQFT is a $\mathbb{Z}_k$ gauge theory stacked with a $\mathbb{Z}_N^{(1)}$ one-form SPT as in \eqref{eq:TQFT_ansatz}, where the parameter $p$ is given by the solution of \eqref{eq:solution}. Using \eqref{eq:solution}, it is straightforward to check that the theory as a spin TQFT is invariant under gauging the $\mathbb{Z}_N^{(1)}$ one-form symmetry and $e^{i\Omega(X)}=1$ on simply-connected smooth spin 4-manifold $X$.

\section{Lattice Examples}\label{sec:lattice}
As an application of the analysis of the previous section, let us study (3+1)d $\mathbb{Z}_{N}$ lattice gauge theory in the Villain formulation \cite{Villain:1974ir}, which is defined on a 4d Euclidean hyper-cube lattice. The action is 
\begin{equation}\label{villain}
    S_{\mathcal{Q}} = \frac{1}{2g^2} \sum_{\textrm{plaquette}}\left(\Delta m^{(1)} - N n^{(2)}\right)^2~,
\end{equation}
where $m^{(1)}$ is the integer one-form gauge field on each link, while $n^{(2)}$ is the integer two-form gauge field on each plaquette. There is an integer gauge symmetry 
\ie
&m^{(1)}\rightarrow m^{(1)}+\Delta k^{(0)}+Nk^{(1)}~,
\\
&n^{(2)}\rightarrow n^{(2)}+\Delta k^{(1)}~,
\fe
that effectively makes the one-form gauge field a $\mathbb{Z}_N$ gauge field. This theory has an electric $\mathbb{Z}_{N}^{(1)}$ one-form global symmetry that shifts $m^{(1)}$ by a flat integer gauge field. By performing a Poisson resummation on $n^{(2)}$ in \eqref{villain}, we obtain a dual description of the theory in terms of an integer gauge field $\hat{n}^{(2)}$ on the plaquettes of the dual lattice. After introducing the Stueckelberg fields $\widetilde m^{(1)}$ and $\widetilde{n}^{(2)}$, we obtain
\ie\label{dual_villain}
S_{S\mathcal{Q}} =&\, \frac{2 \pi i}{N} \sum_{\textrm{link}}m^{(1)} \Delta \widetilde{n}^{(2)}+
\\
&\,
\frac{4 \pi^2 g^2}{2 N^2}  \sum_{\textrm{plaquette}}\left(\Delta \widetilde m^{(1)}-N\widetilde{n}^{(2)}-\hat{n}^{(2)}\right)^2~. 
\fe
By comparing the second term in \eqref{dual_villain} with \eqref{villain}, we see that the duality maps the theory at coupling $g^2$ to the one at coupling $N^2/4\pi^2g^2$ with the electric $\mathbb{Z}_N^{(1)}$ one-form symmetry being gauged \cite{Wegner:1984qt,PhysRevD.17.2637,PhysRevD.19.3715,PhysRevD.19.3698,Ukawa:1979yv,Savit:1979ny}.
At the self-dual coupling $g^2_{\ast} = N/2\pi$, this implies that the theory is invariant under gauging the electric $\mathbb{Z}_{N}^{(1)}$ one-form symmetry and therefore has a non-invertible duality symmetry \cite{Choi:2021kmx}. Consequently, we can apply theorem \ref{thm:TQFT} to infer that the $\mathbb{Z}_N$ lattice gauge theory at the self-dual coupling should either be gapless or spontaneously break the duality symmetry unless $N$ is of the form $N=k^2 \ell$, with $-1$ being a quadratic residue modulo $\ell$. All values of $N\leq 60$ where duality invariant SPTs or TQFTs are not ruled out are listed in Table \ref{ns}.

\begin{figure}[htb!]
    \includegraphics[width = 0.4\textwidth]{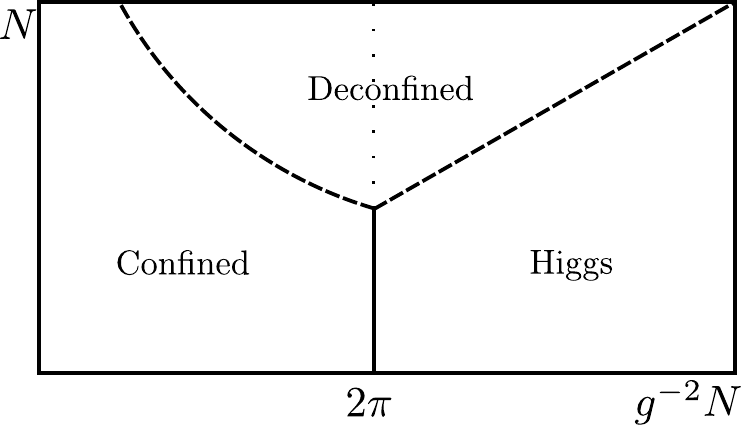}
    \caption{Schematic Phase Diagram of $\mathbb{Z}_{N}$ Lattice Gauge theory. Solid and dashed lines represent first and second order transitions respectively. The vertical line at coupling $g^{2}_{\ast}=N/2\pi$ is duality invariant. 
    At weak coupling, the theory is in the Higgs phase \textit{i.e.}\ it flows to a $\mathbb{Z}_N$ TQFT in the IR and at strong coupling, the theory is in the confining phase \textit{i.e.}\ the vacuum is trivial. For $N\leq 4$, the Higgs and confining phases are separated by a first order phase transition at the self-dual coupling $g^2_{\ast}=N/2\pi$ where the duality symmetry is spontaneously broken. For $N\geq5$, the two phases are separated by an intermediate gapless coulomb phase which flows to a Maxwell theory and at the self-dual coupling $g^2_{\ast}=2\pi/N$ the IR Maxwell theory has $e^2=2\pi/N$.   }
    \label{phases}
\end{figure}

The schematic phase diagram of the $\mathbb{Z}_N$ lattice gauge theory as a function of $N$ and coupling $g$ is summarized in Figure \ref{phases}, with the vertical line at $g_{\ast}^{-2}N= 2\pi$ demarcating the self-dual coupling. This phase diagram is supported by Monte-Carlo simulations \cite{Creutz1979}. When $N\leq 4$, at the self-dual coupling, the duality symmetry is spontaneously broken leading to two local vacua: one is trivial and preserves the $\mathbb{Z}_N^{(1)}$ electric one-form symmetry, and the other one supports a deconfined $\mathbb{Z}_N$ TQFT that spontaneously breaks the electric $\mathbb{Z}_N^{(1)}$ one-form symmetry.  See appendix \ref{app:SSB} for more discussion regarding the spontaneous breaking of non-invertible symmetries. On the other hand, when $N\geq 5$, the duality symmetry is preserved and the theory flows to the gapless Maxwell theory ($U(1)$ gauge theory) at a particular coupling
\ie
S_{\text{Maxwell}}=\frac{1}{2e^2}\int F\wedge\star F~, \quad e^2=\frac{2\pi}{N}~.
\fe
The coupling above is fixed by matching the duality symmetry across the RG flow. The Maxwell theory at $e^2=2\pi/N$ is invariant under gauging the $\mathbb{Z}_N^{(1)}$ subgroup of the $U(1)^{(1)}$ electric one-form symmetry thanks to the electromagnetic duality \cite{Choi:2021kmx}. Thus, the advocated phase diagram is consistent with the constraints stated in theorem \ref{thm:TQFT}.

More generally, one can consider a broader class of lattice gauge theories by introducing a $\theta$ term on the lattice. One concrete realization is the Cardy-Rabinovici model \cite{Cardy1982,Cardy:1981fd} or its generalization \cite{Sulejmanpasic:2019ytl,Anosova:2022cjm}, which is a $U(1)$ lattice gauge theory that couples to charge $N$ electric matter and charge $1$ magnetic matter.\footnote{For simplicity, we consider only a version of the Cardy-Rabinovici model where both bosonic and fermionic matter is present. Otherwise, the duality group in general differs from $SL(2,\mathbb{Z})$ \cite{Metlitski:2015yqa}.} The lattice model is parameterized by a complex coupling $\tau$. We will work with a formal continuum description of the Cardy-Rabinovici model given in \cite{Honda2020}.\footnote{ We can also work with the self-dual lattice models constructed in \cite{Sulejmanpasic:2019ytl,Anosova:2022cjm} which have the $SL(2,\mathbb{Z})$ duality as an exact duality on the lattice.} In the formal continuum description, the Cardy-Rabinovici model is described as a continuum $U(1)$ gauge theory 
\ie
\frac{1}{2e^2} \int F\wedge\star F+\frac{i N\theta}{8\pi^2} \int F\wedge F~,
\fe
at coupling $\tau=2\pi i/ Ne^{2} + \theta/ 2\pi$ with a sum over insertions of charge $N$ Wilson lines and charge 1 't Hooft lines. The sum is weighted by the matter action. In the continuum approximation, this theory  has a $SL(2,\mathbb{Z})$ duality generated by the $\mathbb{S}$ and $\mathbb{T}$ duality transformation. The $\mathbb{S}$ duality maps $\tau\rightarrow-1/\tau$ and swaps the electric and magnetic matter. Since the charges of electric and magnetic matter are different, the duality maps the theory at coupling $\tau$ to the one at $-1/\tau$ with the $\mathbb{Z}_N^{(1)}$ electric one-form symmetry gauged. The $\mathbb{T}$ duality maps $\tau\rightarrow \tau+1$ and turns the magnetic matter to dyonic matter with magnetic charge 1 and electric charge $N$.

The model has a duality symmetry at $\tau=i$, which leads to the same constraint as in the (3+1)d $\mathbb{Z}_N$ lattice gauge theory at the self-dual coupling.

More interestingly, at $\tau_{\ast} = e^{\pi i/3}$, because of the $\mathbb{S}$ duality the model is invariant under the $ST^{-1}$ gauging of the electric $\mathbb{Z}_N^{(1)}$ one-form symmetry and therefore has a triality symmetry.\footnote{If the theory $\mathcal{Q}$ is invariant under $ST^{-1}$ operation, the theory $T^{-1}\mathcal{Q}$ is invariant under $T^{-1}S$ operation.} According to theorem \ref{thm:TQFT_triality}, the model either is gapless or spontaneously breaks the triality symmetry unless $N=k^2\ell$ for some integers $k$ and $\ell$ such that there exists a solution to $
p(p+1)+1=0\text{ mod }\ell$.  Even if $N$ obeys the condition, we can still exclude the possibility of triality invariant TQFTs using the mixed triality-gravity anomaly. In \cite{Hayashi2022}, Hayashi and Tanizaki showed that on a spin manifold $X$, the model at the complex coupling $\tau_{\ast} = e^{i \pi/3}$ has a mixed triality-gravity anomaly given by 
\begin{equation}
    Z_{ST^{-1}\mathcal{Q}}[X] = Z_{\mathcal{Q}}[X] \exp{\bigg[- \frac{i \pi}{3} \sigma(X)\bigg]}
    ~. 
\end{equation}
In particular, we can pick $X$ to be the simply-connected smooth spin 4-manifold $K3$ whose  
signature is $-16$. We then have $Z_{ST^{-1}\mathcal{Q}}[K3]/Z_{\mathcal{Q}}[K3]=e^{-{2 \pi i}/{3} }$.
However, theorem \ref{thm:TQFT_triality} states that any triality invariant TQFT with a unique local vacuum has $Z_{ST^{-1}\mathcal{Q}}[X]/Z_{\mathcal{Q}}[X]=1$ on simply-connected smooth spin 4 manifold $X$.
Therefore, we can deduce that the theory at  $\tau_{\ast} = e^{i \pi/3}$ cannot flow to a triality symmetry preserving TQFT. Note that this is a stronger result than that of \cite{Hayashi2022}, wherein it is shown that the theory cannot be in a SPT phase. 

A heuristic calculation of the free-energy of dyons \cite{Cardy1982} suggests the following scenario for the Cardy-Rabinovici model at $\tau_{\ast}=e^{\pi i/3}$. When $N$ is small, the triality symmetry is spontaneously broken leading to three vacua: a Higgs vacuum (a $\mathbb{Z}_N$ TQFT that spontaenously breaks the $\mathbb{Z}_N^{(1)}$ one-form symmetry), a confined vacuum (a trivial vacuum that preserves the $\mathbb{Z}_N^{(1)}$ one-form symmetry) and a oblique confined vacuum (a $\mathbb{Z}_N^{(1)}$ one-form SPT) \cite{Moy:2022ztf}. The triality symmetry permutes these three vacua. When $N$ is large, the theory flows to the triality invariant gapless Maxwell theory. By matching the triality symmetry, we can fix the coupling of the Maxwell theory to be the triality symmetric point \cite{Choi:2022zal}
\ie
S_{\text{Maxwell}}=\frac{N}{4\pi} \frac{\sqrt{3}}{2}\int F\wedge\star F+\frac{N}{4\pi} \frac{i}{2} \int F\wedge F~.
\fe
These two scenarios are both consistent with the constraint stated in theorem \ref{thm:TQFT_triality}.

\let\oldaddcontentsline\addcontentsline% Store \addcontentsline
\renewcommand{\addcontentsline}[3]{}% Make \addcontentsline a no-op
\section*{Acknowledgements}
\let\addcontentsline\oldaddcontentsline% Restore \addcontentsline

We thank K. Ohmori, and R. Thorngren for discussions. AA is supported by Yoichiro Nambu Graduate Fellowship courtesy of Department of Physics, University of Chicago. CC is supported by the US Department of Energy DE-SC0021432 and the Simons Collaboration on Global Categorical Symmetries. HTL is supported in part
by a Croucher fellowship from the Croucher Foundation, the Packard Foundation and the
Center for Theoretical Physics at MIT.

\onecolumngrid

\appendix
\section{Gauss Sums of Intersection Forms}\label{secgauss}

Let $X$ be a smooth simply connected 4-manifold, with the intersection form $I: H_{2}(X;\mathbb{Z}) \times H_{2}(X;\mathbb{Z}) \to \mathbb{Z}$. The intersection form $I$ is a bi-linear symmetric form. We are interested in computing its normalized Gauss sum given by
\begin{equation}\label{eq:Gauss_sum_comp}
    G(I,p,N)=\frac{1}{N^{\textrm{rk}(I)/2}}\sum_{x_{i}=1}^{N} \exp{\bigg(\frac{2 \pi i p}{2N} x_{i}I_{ij}x_{j}\bigg)}~,
\end{equation}
where $\textrm{rk}(I)$ is the rank of $I$. The normalized Gauss sum can be interpreted as the partition function of a $\mathbb{Z}_N$ two-form gauge theory when $pN$ is even (see appendix \ref{app:partitionfunction} for more details). In this appendix, we will focus on the case that has even $pN$ and $p$ co-prime to $N$.

Note that the intersection form $I$ may be definite or indefinite. Let $E_{8}$ be the intersection form of the $E_{8}$ manifold and $H$ be the intersection form of $S^{2} \times S^{2}$, then by combining 
Donaldson's theorem with Hasse-Minkowski theorem we have the following result \cite{Scorpan}
\begin{theorem}
The only bi-linear symmetric forms that can be realized as intersection forms of a smooth simply connected 4-manifold are 
\begin{equation}
    \oplus ~a [+1] ~,~ \oplus  ~b [-1] ~,~ \oplus \pm m E_{8} \oplus n H~;
\end{equation}
where $a,b,m,n$ are positive integers. 
\end{theorem}
By the theorem above, the task of computing the Gauss sum \eqref{eq:Gauss_sum_comp} for a general intersection form reduces to computing the Gauss sum for $[+1], [-1], E_{8}$ and $H$. 

To prepare for the calculation below, define
\begin{align}
\varepsilon_{N} =
  \begin{cases}
    1 ~&\textrm{if}~ N \equiv 1 \mod{4} \\
    i ~&\textrm{if}~ N \equiv 3 \mod{4}
  \end{cases}~.
\end{align}
Let us also review the Jacobi symbol $(\frac{a}{n})$, which is defined for any integer $a$ and any positive odd integer $n$. The Jacobi symbol factorizes into the product of the Legendre symbols
\ie
\left(\frac{a}{n}\right)=\left(\frac{a}{s_1}\right)^{r_1}\left(\frac{a}{s_2}\right)^{r_2}\cdots\left(\frac{a}{s_k}\right)^{r_k}~,
\fe
where $n=s_1^{r_1}s_2^{r_2}\cdots s_k^{r_k}$
is the prime factorization of $n$. The Legendre symbols is defined for any integer $a$ and any odd prime $s$ as
\ie
\left(\frac{a}{s}\right)=\begin{cases}
	0\quad &\text{if }a=0\text{ mod }s
	\\
	1\quad &\text{if }a\neq0\text{ mod } s\text{ and } a=x^2\text{ mod } s\text{ for some integer }x
	\\
	-1\quad &\text{if }a\neq 0\text{ mod }s\text{ and there is no such $x$}
\end{cases}
\fe

The Gauss sum for $I=[+1]$ is \cite{Nagell1981}
\begin{equation}\label{plusone}
    G([+1],p,N)=\begin{cases}
    \varepsilon_{N} \left(\frac{p/2}{N}\right)  &\text{odd } N
    \\
    e^{ \frac{\pi i }{4}}\varepsilon_{ p}^{-1}\left(\frac{ 2N}{p}\right) &\text{even }N
    \end{cases}~.
\end{equation}
It is valued in 4th root of unity when $N$ is odd and 8th root of unity when $N$ is even.
Taking its complex conjugate we obtain the Gauss sum for $I=[-1]$.
For $H$, the Gauss sum is unity 
\begin{equation}\label{H}
    G(H,p,N)= \frac{1}{N}\sum_{x_{1}, x_{2}=1}^{N} \exp{\bigg(\frac{2 \pi i p}{2N} 2 x_{1}x_{2}\bigg)} = 1~.
\end{equation}
For $E_8$, the  Gauss sum is also unity \cite{Hayashi2022}
\ie
G(E_{8},p,N) = 1~. 
\fe
Since $[+1]$, $[-1]$, $H$ and $E_8$ have signature $+1$, $-1$, $0$ and $+8$ respectively, the normalized Gauss sum for the intersection form $I$ of a smooth simply connected 4-manifold $X$ can be summarized a function of the signature $\sigma(X)$ of the 4-manifold
\ie \label{bigform}
G(I,p,N)=\begin{cases}
    \left[\varepsilon_{N} \left(\frac{p/2}{N}\right)\right]^{\sigma(X)}  &\text{odd } N
    \\
    \left[e^{ \frac{\pi i }{4}}\varepsilon_{ p}^{-1}\left(\frac{ 2N}{p}\right)\right]^{\sigma(X)} &\text{even }N
    \end{cases}~.
\fe
Since the normalized Gauss sum is valued in 8th root of unity and by Rokhlin's theorem the signature of the intersection form $I$ of a spin manifold is divisible by 16, the Gauss sum is always unity on a spin-manifold.

We now study two special cases. First, consider the case when $p^2 = -1 \text{ mod }{N}$ and $N$ is odd. This situation arises only if all the prime factors of $N$ are 1 modulo 4, and thus $\varepsilon_N = 1$. Since $p/2 = \left[(2)^{-1}_N(p+1)\right]^2$ mod $N$, we have $\left(\frac{p/2}{N}\right)=1$. Combining these two facts, we find that the normalized Gauss sum is unity:
\begin{equation}\label{gausstrivial}
    G(I,p,N)=1 ~\text{when}~p^2 \equiv -1 \text{ mod }{N}\text{ and } N\text{ is odd} ~.
\end{equation}
Next, consider the case when $p(p+1)=-1$ mod $N$. This is possible only when $N$ is odd. Using the following properties of the Jacobi symbol
\ie
\left(\frac{ab}{n}\right)=\left(\frac{a}{n}\right)\left(\frac{b}{n}\right)~,\quad \left(\frac{a}{n}\right)^2=1~,\quad \left(\frac{1}{n}\right)=1~,
\fe
we find that
\ie
\left(\frac{p/2}{N}\right)=\left(\frac{2}{N}\right)\left(\frac{p}{N}\right)=\left(\frac{2}{N}\right)\left(\frac{(p+1)^2}{N}\right)=\left(\frac{2}{N}\right)~.
\fe
Therefore,
\begin{equation}\label{eq:triality_Gauss_sum}
    G(I,p,N) = \left[\varepsilon_{N}\bigg(\frac{2}{N}\bigg)\right]^{\sigma(X)}\text{ when $p(p+1)=-1$ mod $N$}~.
\end{equation}

\section{Partition Function of $\mathbb{Z}_N$ 2-form Gauge Theory}\label{app:partitionfunction}
In this appendix, we compute the partition function of a (3+1)d $\mathbb{Z}_N$ 2-form gauge theory in the presence of the background gauge field $B$ for the $\mathbb{Z}_N^{(1)}$ one-form symmetry. For simplicity, we assume the underlying manifold $X$ is simply connected and thus $H^2(X,\mathbb{Z})$ has no torsion classes. 

The partition functions of a (3+1)d $\mathbb{Z}_N$ 2-form gauge theory is
\ie\label{eq:even_partition_function}
Z[B]=\frac{1}{\sqrt{|H^2(X,\mathbb{Z}_N)|}}\sum_{b\in H^2(X,\mathbb{Z}_N)}\exp\left[2\pi i\int_X\left(\frac{p}{2N} \mathcal{P}(b)+\frac{1}{N}b\cup B\right)\right]~.
\fe
When $N$ is even, $p\sim p+2N$ and $\mathcal{P}(b):H^2(X,\mathbb{Z}_N)\rightarrow H^4(X,\mathbb{Z}_{2N})$ is the Pontryagin square map. When $N$ is odd, $p\sim p+2N\in 2\mathbb{Z}$ and $\mathcal{P}(b)=b\cup b:H^2(X,\mathbb{Z}_N)\rightarrow H^4(X,\mathbb{Z}_{N})$. In both cases $pN$ is even.
Let us define 
\ie
L\equiv\text{gcd}(p,N),\quad q\equiv \frac{p}{L},\quad K\equiv\frac{N}{L},\quad J\equiv\frac{qKL}{2}~.
\fe
If $q$ and $K$ are both odd, $L$ has to be even because $pN=qKL^2$ is even. Thus, $J$ is always an integer.
To evaluate the partition function, we split $b$ into a $\mathbb{Z}_L$ cochain $b_0$ and $\mathbb{Z}_K$ cocycle $b_1$:
\ie\label{eq:split}
b=Kb_0+b_1~.
\fe
The gauge symmetries are
\ie
b_0\rightarrow b_0+\delta \alpha_0+L\beta_0-\beta_1~,
\quad b_1\rightarrow b_1+\delta \alpha_1+K\beta_1~.
\fe
The cocycle conditions are
\ie
\delta b_1 = 0 \text{ mod }K~,
\quad \delta b_0 = -{\delta b_1}/{K} \text{ mod }L \equiv -\text{Bock}(b_1)  \text{ mod }L~;
\fe
where $\text{Bock}(b_1)$ is the Bockstein homomorphism applied to the cocycle $b_1$. The second cocycle condition ensures that $\delta b=\delta(Kb_0+b_1)=0\text{ mod } N$. Now substituting \eqref{eq:split} into \eqref{eq:even_partition_function}, we obtain
\ie
{Z}[B]=
&\frac{1}{\sqrt{|H^2(X,\mathbb{Z}_N)|}}\times
\\
&\sum_{b_1\in H^2(X,\mathbb{Z}_K)}\exp\left[2\pi i\int_X\left(\frac{q}{2K}\mathcal{P}(b_1)+\frac{1}{N}b_1\cup B\right)\right]\sum_{\substack{b_0\in C^2(X,\mathbb{Z}_L)\\ \delta b_0=-\text{Bock}(b_1)}}\exp\left[2\pi i\int_X\left(\frac{ qK}{2}\mathcal{P}(b_0)+\frac{1}{L}b_0\cup B\right)\right]~.
\fe
The second sum can be simplfied using the Wu formula, $\mathcal{P}(b)=b\cup \nu_2(X)$ mod $2$, where $\nu_2(X)=w_2(X)+w_1(X)\cup w_1(X)$ is the second Wu class of the underlying manifold $X$ and $w_i(X)$ is the $i$th Stiefel-Whitney class.
Summing over $b_0$ then gives
\ie
{Z}[B]=&\frac{|H^2(X,\mathbb{Z}_L)|}{\sqrt{|H^2(X,\mathbb{Z}_N)|}}\times
\\
&\sum_{b_1\in H^2(X,\mathbb{Z}_K)}\exp\left[2\pi i\int_X\left(\frac{q(1-K)}{2K}\mathcal{P}(b_1)+\frac{1}{K}b_1\cup \frac{1}{L}\left(B+J\nu_2(X)\right)\right)\right]\delta\left((B+J\nu_2(X))\text{ mod }L\right)~.
\fe
We reorganized the exponents using the Wu formula so that each term is independently well-defined. The first term is well-defined because $qK(1-K)$ is always even. The second term is well-defined when the delta function is non-zero.
We can complete the square in the exponent by a change of variable
\ie
\tilde b_1 = b_1+(1+K)\left(q\right)^{-1}_{\gamma(K)K}\frac{B+J\nu_2}{L}~,
\fe
where the notation $(\alpha)_\beta^{-1}$ denotes the inverse of $\alpha$ in the ring $\mathbb{Z}_\beta$ for co-prime integers $\alpha$ and $\beta$, and
\ie
\gamma(K) = \begin{cases}
    1,\quad & \text{odd }K
    \\
    2,\quad & \text{even }K
\end{cases}.
\fe
When $q=0$, we define $(q)^{-1}_{\gamma(K)K}=0$.
The exponents in the sum now becomes
\ie
\exp\left[2\pi i\int_X\left(\frac{ q(1-K)}{2K}\mathcal{P}(\tilde{b}_1)-\frac{(1-K)\left(q\right)^{-1}_{\gamma(K)K}}{2K}\mathcal{P}\left(\frac{B+J\nu_2}{L}\right)\right)\right]~;
\fe
The remaining sum over $\tilde b_1$ is related to the normalized Gauss sum for the intersection form $I$ via
\begin{equation}
    \sum_{\tilde b_1\in H^2(X,\mathbb{Z}_K)}\exp\left[2\pi i\int_X\frac{ q(1-K)}{2K}\mathcal{P}(\tilde b_1)\right]=\sqrt{|H^2(X,\mathbb{Z}_K)|}G(I,q(1-K),K)~.
\end{equation}
Summing over $b_1$ then leads to
\ie\label{eq:partition_function_ZN}
{Z}[B]=\sqrt{|H^2(X,\mathbb{Z}_L)|}G(I,q(1-K),K) \exp\left[-2\pi i\int_X\frac{(1-K)\left(q\right)^{-1}_{\gamma(K)K}}{2K}\mathcal{P}\left(\frac{B+J\nu_2}{L}\right)\right]\delta\left((B+J\nu_2)\text{ mod }L\right)~.
\fe 

We now discuss two situations when the formula simplifies. When the underlying manifold $X$ has a trivial second Wu class $\nu_2(X)$, \eqref{eq:partition_function_ZN} simplifies to 
\ie\label{eq:partition_function_ZN_even_qK}
{Z}[B]=\sqrt{|H^2(X,\mathbb{Z}_L)|}G(I,q,K)\delta\left(B\text{ mod }L\right)
\exp\left[-2\pi i\int_X\frac{\left(q\right)^{-1}_{K}}{2K}\mathcal{P}\left(\frac{B}{L}\right)\right]
~.
\fe 
In the special case of $L=1$, \eqref{eq:partition_function_ZN} simplifies to
\ie\label{eq:partition_function_ZN_L=1}
{Z}[B]=G(I,p,N)
\exp\left[-\int_X\frac{2\pi i\left(p\right)^{-1}_{\gamma(N)N}}{2N}\mathcal{P}\left({B}\right)\right]
~.
\fe 
When $N$ is odd, we pick an even $(p)^{-1}_N$ for the expression to be well-defined. This can always be achieved by redefining $(p)^{-1}_N$ to $(p)^{-1}_N+N$.

\section{Details of TQFT Argument}\label{app:TQFT}
In this appendix, we provide more details of the proof of theorem \ref{thm:TQFT} and theorem \ref{thm:TQFT_triality}. 

\subsection{Duality Symmetry}
Consider a (3+1)d (spin) TQFT $\mathcal{Q}$ that has a unique vacuum and a $
\mathbb{Z}_N^{(1)}$ one-form symmetry. We will prove that the TQFT cannot be invariant under gauging the $
\mathbb{Z}_N^{(1)}$ one-form symmetry up to a gravitational counterterm $e^{i\Omega(X)}$ unless $N=k^2\ell$ with $k,\ell\in\mathbb{Z}$ and $-1$ is a quadratic residue of $\ell$. Furthermore, if the TQFT is invariant under gauging, the gravitational counterterm $e^{i\Omega(X)}=1$ on every simply-connected smooth spin 4-manifold $X$.

The $
\mathbb{Z}_N^{(1)}$ one-form symmetry is generated by surface operators $U_g$. To detect the one-form charge of a line operator $\mathcal{L}$, we link the surface operators $U_g$ with the line operator $\mathcal{L}$. If the correlation function, where $U_g$ and $\mathcal{L}$ form a Hopf link, differs from the one, where the link is trivial, we say the surface operator $U_g$ links nontrivially with the line operator $\mathcal{L}$. Otherwise, we say $U_g$ and $\mathcal{L}$ link trivially with each other.

In general, the $
\mathbb{Z}_N^{(1)}$ one-form symmetry can act unfaithfully on line operators in the sense that some symmetry operators $U_g$ link trivially with all line operators.\footnote{An unfaithful symmetry operator can still have non-trivial correlation functions associated to other topological invariants, such as intersections, triple linkings and quadruple linkings of surfaces and etc. \cite{Putrov:2016qdo, Hsin:2019fhf}.} The unfaithful one-form symmetry forms a $
\mathbb{Z}_M^{(1)}$ subgroup of the $
\mathbb{Z}_N^{(1)}$ one-form symmetry where $M\in\mathbb{Z}$ and $k=N/M\in\mathbb{Z}$. In this case, all line operators transform with a $k$th root of unity, instead of a $N$th root of unity, under the $
\mathbb{Z}_N^{(1)}$ one-form symmetry. 

We now gauge the $
\mathbb{Z}_N^{(1)}$ one-form symmetry by coupling the theory to a dynamical $\mathbb{Z}_N$ two-form gauge field $b\in H^2(X,\mathbb{Z}_N)$. This leads to a dual $
\hat{\mathbb{Z}}_N^{(1)}$ one-form symmetry, generated by the Wilson surface operator $\hat{U}=\exp({2\pi i \oint b/N})$, in the gauged theory $\mathcal{Q}/\mathbb{Z}_N^{(1)}$. Here, we add a hat to the dual symmetry group to distinguish it from the original symmetry group.

Line operators charged under the $
\mathbb{Z}_N^{(1)}$ one-form symmetry become non-gauge-invariant in the gauged theory $\mathcal{Q}/\mathbb{Z}_N^{(1)}$. To make them gauge invariant, we attach an open Wilson surface operator to them. Recall that all line operators transform only with a $k$th root of unity under the $
\mathbb{Z}_N^{(1)}$ one-form symmetry. Therefore, only the Wilson surface operators generated by $\hat{U}^M=\exp(2\pi i\oint b/k)$ can end on these charged lines. The charged lines are topological, based on the assumption that $\mathcal{Q}$ is a TQFT, so they provide a topological boundary condition to the Wilson surface operators ending on them. Using these topological boundary conditions, we can unlink any Hopf link between the surface operators generated by $\hat{U}^M=\exp(2\pi i\oint b/k)$ and any line operator. Thus, the $
\hat{\mathbb{Z}}_{k}^{(1)}$ subgroup, generated by $\hat{U}^M=\exp(2\pi i\oint b/k)$, of the dual $
\hat{\mathbb{Z}}_N^{(1)}$ one-form symmetry acts unfaithfully in the gauged theory $\mathcal{Q}/\mathbb{Z}_N^{(1)}$. 

The unfaithful subgroup of the dual $\hat{\mathbb{Z}}_N^{(1)}$ one-form symmetry can be larger than $\hat{\mathbb{Z}}_{k}^{(1)}$. If the TQFT is invariant under gauging, the unfaithful subgroup in the gauged theory $\mathcal{Q}/\mathbb{Z}_N^{(1)}$ should be $
\hat{\mathbb{Z}}_{M}^{(1)}$, which should include $
\hat{\mathbb{Z}}_{k}^{(1)}$ as a subgroup. It implies that $N=k^2\ell$ for some integer $\ell$. 

We now derive a constraint on $\ell$. Consider placing the theory $\mathcal{Q}$ on $S^2\times S^2$, which has $H_2(S^2\times S^2,\mathbb{Z})=H^2(S^2\times S^2,\mathbb{Z})=\mathbb{Z}^2$ and an intersection form $H$. We can turn on the background gauge field $(B_1,B_2)\in H^2(S^2\times S^2,\mathbb{Z}_M)=\mathbb{Z}_M^2$ for the unfaithful $
\mathbb{Z}_M^{(1)}$ subgroup one-form symmetry. It amounts to inserting $B_1$ and $B_2$ number of basic $
\mathbb{Z}_{M}^{(1)}$ symmetry operators wrapping around the two $S^2$ respectively. According to Proposition 1 of \cite{Cordova:2019bsd}, any surface operator in a TQFT that links trivially with all line operators admits a topological boundary condition. In our TQFT $\mathcal{Q}$, it implies that the symmetry operators of the unfaithful $
\mathbb{Z}_{M}^{(1)}$ subgroup one-form symmetry can be opened up topologically without changing the correlation functions. On $S^2\times S^2$, opening up the $(B_1,B_2)$ number of symmetry operators and shrinking them leaves us with $B_1B_2$ number of identical local operators $\mathcal{O}$ at the intersections. Since the TQFT has a unique local vacuum, these local operators must be a multiple of the identity operator, i.e.\ $\mathcal{O}=\lambda\mathbf{1}$ and therefore the partition function is $Z_\mathcal{Q}[B_1,B_2]=\lambda^{B_1B_2}Z_\mathcal{Q}[0,0]$. Because of the identification $B_{1,2}\sim B_{1,2}+M$, the coefficient $\lambda$ obeys
\ie
\frac{Z_\mathcal{Q}[B_1+M,B_2]}{Z_\mathcal{Q}[B_1,B_2]}=\lambda^{MB_2}=1~.
\fe
Solving for $\lambda$, we get
\ie\label{eq:Z[B]}
Z_\mathcal{Q}[B_1,B_2]=Z_\mathcal{Q}[0,0]\exp\left[\frac{2\pi i p}{M}B_1B_2\right]~,\quad p\in\mathbb{Z}~.
\fe

The fact that $\mathcal{Q}$ is invariant under gauging the $
\mathbb{Z}_N^{(1)}$ one-form symmetry imposes a constraint on $Z_\mathcal{Q}[B_1,B_2]$. Starting from the theory $\mathcal{Q}$, we can gauge the $
\mathbb{Z}_M^{(1)}$ subgroup one-form symmetry and obtain the theory $\mathcal{Q}/\mathbb{Z}_M^{(1)}$, which has a dual $
\tilde{\mathbb{Z}}_M^{(1)}$ and a $
\tilde{\mathbb{Z}}_k^{(1)}$ one-form symmetry. We put a tilde on the symmetry group of the theory $\mathcal{Q}/\mathbb{Z}_M^{(1)}$. The dual $
\tilde{\mathbb{Z}}_M^{(1)}$ one-form symmetry is generated by the Wilson surface operators of the $
\mathbb{Z}_M^{(1)}$ two-form gauge field. Turning on the background gauge field $ (C_1,C_2)\in H^2(S^2\times S^2,\mathbb{Z}_M)$ for the $
\tilde{\mathbb{Z}}_\ell^{(1)}$ subgroup of the $
\tilde{\mathbb{Z}}_M^{(1)}$ one-form symmetry leads to the following partition function for the $\mathcal{Q}/\mathbb{Z}_M^{(1)}$ theory on $S^2\times S^2$ that can be computed with the formula for partition functions \eqref{eq:Z[B]} and \eqref{eq:partition_function_ZN_even_qK}
\ie\label{eq:ZM1}
Z_{\mathcal{Q}/\mathbb{Z}_M^{(1)}}[C_1,C_2]&=\frac{1}{M}\sum_{b_{1,2}=1}^M Z_{\mathcal{Q}}[b_1,b_2]\exp\left[\frac{2\pi i}{\ell}(b_1C_2+b_2C_1)\right]
\\
&=\frac{1}{M}\sum_{b_{1,2}=1}^{k\ell} 
Z_\mathcal{Q}[0,0]\exp\left[\frac{2\pi i p}{k\ell}b_1b_2+\frac{2\pi i}{\ell}(b_1C_2+b_2C_1)\right]
\\
&=Z_\mathcal{Q}[0,0]\times L\exp\left[-\frac{2\pi i (q)^{-1}_K}{K}\frac{kC_1}{L}\frac{kC_2}{L}\right]\delta(k C_{1,2}\text{ mod }L)~.
\fe
where $L=\text{gcd}(p,k\ell)$, $K=k\ell/L$, $q=p/L$. Note that the Gauss sum equals unity for the intersection pairing on $S^2\times S^2$ as shown in the appendix on Gauss sums. 
On the other hand, we can also start from the gauged theory $\mathcal{Q}/\mathbb{Z}_N^{(1)}$ and gauge the $
\hat{\mathbb{Z}}_k^{(1)}$ subgroup of the $
\hat{\mathbb{Z}}_N^{(1)}$ one-form symmetry to obtain the theory $\mathcal{Q}/\mathbb{Z}_M^{(1)}$. The $
\tilde{\mathbb{Z}}_k^{(1)}$ one-form symmetry of $\mathcal{Q}/\mathbb{Z}_M^{(1)}$ is the dual symmetry generated by the Wilson surface operators of the $
\hat{\mathbb{Z}}_k^{(1)}$ two-form gauge field. The $
\hat{\mathbb{Z}}_M^{(1)}/\hat{\mathbb{Z}}_k^{(1)}=\mathbb{Z}_\ell^{(1)}$ part of the $
\hat{\mathbb{Z}}_M^{(1)}$ one-form symmetry of $\mathcal{Q}/\mathbb{Z}_N^{(1)}$ becomes the $
\tilde{\mathbb{Z}}_\ell^{(1)}\subset 
\tilde{\mathbb{Z}}_M^{(1)}$ one-form symmetry of $\mathcal{Q}/\mathbb{Z}_M^{(1)}$. We then have another expression for $Z_{\mathcal{Q}/\mathbb{Z}_M^{(1)}}[C]$
\ie\label{eq:ZM2_before}
Z_{\mathcal{Q}/\mathbb{Z}_M^{(1)}}[C_1,C_2]=\frac{1}{\sqrt{k}}\sum_{b_{1,2}=1}^k Z_{\mathcal{Q}/\mathbb{Z}_N^{(1)}}[\ell b_1+C_1,\ell b_2+C_2]~.
\fe

Recall that the partition functions before and after gauging are identical up to a gravitational counterterm  $e^{i\Omega(X)}$ which is a pure phase. On $S^2\times S^2$, this gives
\ie
Z_{\mathcal{Q}}[B_1,B_2]=Z_{\mathcal{Q}/\mathbb{Z}_N^{(1)}}[B_1,B_2]e^{-i\Omega(S^2\times S^2)}~.
\fe
Together with \eqref{eq:Z[B]} and the formula \eqref{eq:partition_function_ZN_even_qK}, we obtain
\ie\label{eq:ZM2}
Z_{\mathcal{Q}/\mathbb{Z}_M^{(1)}}[C_1,C_2]&=\frac{1}{k}\sum_{b_{1,2}=1}^k 
Z_\mathcal{Q}[0,0]e^{i\Omega(S^2\times S^2)}\exp\left[\frac{2\pi i p\ell}{k}b_1b_2+\frac{2\pi ip}{k}(b_1C_2+b_2C_1)+\frac{2\pi i p}{k\ell}C_1C_2\right]
\\
&=Z_\mathcal{Q}[0,0]e^{i\Omega(S^2\times S^2)}\times L\exp\left[\frac{2\pi ip}{k\ell}C_1C_2-\frac{2\pi i (\tilde q)^{-1}_{\tilde K}}{\tilde K}\frac{pC_1}{\tilde L}\frac{pC_2}{\tilde L}\right]\delta(p C_{1,2}\text{ mod }\tilde L)~,
\fe
where $\tilde L=\text{gcd}(p\ell,k)$, $\tilde K=k/\tilde L$, $\tilde q= p\ell/\tilde L$.

On $S^2\times S^2$, the partition function $Z_{\mathcal{Q}}[0,0]$ is positive \cite{Cordova:2019bsd}.\footnote{This is because $S^{2}\times S^{2}$ is the double of an open four-manifold $\chi$ which may be embedded inside $S^{4}$.  In particular this means that the partition function of $\mathcal{Q}$ on $S^{4}$ which is necessarily non-zero by unitarity, may be viewed as the inner-product of the state defined by $\chi$ and the state defined by the complement of $\chi$. Therefore $\chi$ defines a non-zero state and the partition function on $S^{2}\times S^{2}$ is the norm of a non-zero vector and so positive.} Hence, equating \eqref{eq:ZM1} and \eqref{eq:ZM2}, we get the following constraints. First of all, to match the magnitudes of the two partition functions, we have
\ie\label{eq:LtildeL}
L\delta\left(kC_{1,2}\text{ mod }L\right)=\tilde L\delta\left(pC_{1,2}\text{ mod }\tilde L\right)~,\quad L\equiv\text{gcd}(p,k\ell)~,\quad \tilde L\equiv\text{gcd}(p\ell,k)~.
\fe
It implies that $L=\tilde L=\text{gcd}(p,k)$ and therefore $\ell$, $\hat p\equiv p/\text{gcd}(p,k)$, $\hat k\equiv k/\text{gcd}(p,k)$ are all co-prime to each other.
Once these conditions are obeyed, the delta functions in \eqref{eq:ZM1} and \eqref{eq:ZM2} both become trivial. Next, we match the phases of the two partition functions. Since $e^{i\Omega(S^2\times S^2)}$ is independent of $(C_1,C_2)$, we have
\ie
e^{i\Omega(S^2\times S^2)}=1~.
\fe
Matching the remaining phases that depend on $(C_1,C_2)$ leads to the constraint
\ie\label{eq:ref1}
-\frac{(\hat p)^{-1}_{\hat k\ell}}{\hat k\ell}\hat k^2=\frac{\hat p}{\hat k\ell}-\frac{(\hat p \ell)^{-1}_{\hat k}}{\hat k}\hat p^2\text{ mod }1
\fe
We can multiply the equation by $\hat k \ell$ and treat it as an equation modulo $\ell$.\footnote{If we multiply the equation \eqref{eq:ref1} by $\hat k \ell$ and treat it as an equation modulo $\hat k$, we get a trivial equality. Since $\hat k$ and $\ell$ are co-prime, the equality \eqref{eq:ref1} is equivalent to \eqref{ref2}.} The second term on the right-hand side dropped out and the constraint simplifies to
\ie\label{ref2}
(\hat p)^{-1}_{\hat k\ell}\hat k^2+{\hat p}=0\text{ mod }\ell
\fe
If we further multiply the equality by $(\hat p)^{-1}_{\hat k\ell}$ and still treat the equality as an equality modulo $\ell$, we get 
\ie
(\hat k(\hat p)^{-1}_{\hat k\ell})^2+1=0\text{ mod }\ell~.
\fe
The constraint implies that $-1$ is a quadratic residue of $\ell$. The condition holds if and only if every odd prime factor $x_i$ in the prime factorization of $\ell$
\ie
\ell = 2^{y_0} x_1^{y_1}x_3^{y_3}\cdots x_m^{y_m}~,
\fe
is 1 modulo 4 and the power of $2$ in the prime factorization is $y_0=0,1$. This completes the first part of the proof.

The foregoing derivation remains unmodified if we replace the spacetime manifold $S^2 \times S^2$ by any other simply-connected smooth spin 4-manifold $X$. On such manifolds, by opening up  the unfaithful $\mathbb{Z}_M^{(1)}$ symmetry operator and shrinking them to the intersection points, we can fix the partition function to
\ie\label{eq:spin_manifold_partition}
Z_\mathcal{Q}[B]=Z_\mathcal{Q}[0]\exp\left[\frac{2\pi ip}{M}\frac{B^TIB}{2}\right]~, \quad p\in\mathbb{Z}~,
\fe
where $B\in H^2(X,\mathbb{Z}_M)=\mathbb{Z}_M^n$ is the background gauge field for 
the unfaithful $\mathbb{Z}_M^{(1)}$ one-form symmetry and $I$ is the intersection form of the 4-manifold $X$. Here, $p$ is an integer because  the intersection form of a spin manifold is even. Note that $Z_\mathcal{Q}[0]$ is positive on any simply-connected smooth spin 4-manifold
\cite{Cordova:2019jqi}. Generalizing \eqref{eq:ZM1} and \eqref{eq:ZM2}, we have the equality
\ie\label{eq:spin_manifold_equality}
\frac{1}{M^{\text{rk}(I)/2}}\sum_{b=1}^MZ_\mathcal{Q}[b]\exp\left[\frac{2\pi i}{\ell} b^T I C\right]=\frac{1}{k^{\text{rk}(I)/2}}\sum_{b=1}^kZ_{\mathcal{Q}/\mathbb{Z}_N^{(1)}}[\ell b+C]~.
\fe
Substiting \eqref{eq:spin_manifold_partition} and \ie
Z_{\mathcal{Q}}[B]=Z_{\mathcal{Q}/\mathbb{Z}_N^{(1)}}[B]e^{-i\Omega(X)}
\fe
into \eqref{eq:spin_manifold_equality} leads to the same constraint on $\ell$ as before and since the Gauss sum for the intersection forms of $X$ is unity, we have $e^{i\Omega(X)}=1$ on every simply-connected smooth spin 4-manifolds. This concludes the proof. Since the proof only uses spin 4-manifolds, which have a trivial second Stiefel-Whitney class, we cannot determine whether the TQFT is a spin TQFT or a non-spin TQFT. 

\subsection{Triality Symmetry and More General Non-invertible Symmetry}
We now generalize the constraint on duality invariant TQFTs to TQFTs that preserves more general non-invertible symmetry including the triality symmetry. Consider a (3+1)d (spin) TQFT $\mathcal{Q}$ that has a unique local vacuum and a $
\mathbb{Z}_N^{(1)}$ one-form symmetry. We will prove that the TQFT cannot be invariant under $T^{-p'}S$ gauging the $
\mathbb{Z}_N^{(1)}$ one-form symmetry upto a gravitational counterterm $e^{i\Omega(X)}$ unless $N=k^2\ell$ with $k,\ell\in\mathbb{Z}$ and there exists a solution $p\in\mathbb{Z}$ such that
\ie
 p( p+p')+1=0\text{ mod }\ell~.
\fe
Furthermore, if the TQFT is invariant under 
the $T^{-p'}S$ gauging, the gravitational counterterm $e^{i\Omega(X)}=1$ on every simply-connected smooth spin 4-manifold $X$.
When $N$ is even and $p'=\pm 1$ or when $N$ is odd and $p'=\pm (1+N)$, the non-invertible defects constructed by half-gauging the theory leads to the triality defects.

The proof for these more general $T^{-p'}S$ symmetries follows the same reasoning as in the duality case.  We will directly work on simply connected smooth spin manifolds $X$.
\eqref{eq:spin_manifold_partition} and \eqref{eq:spin_manifold_equality} still holds but now due to the invariance under $T^{-p'}S$ gauging we have 
\ie
Z_{\mathcal{Q}/\mathbb{Z}_{N}^{(1)}}[B]=Z_{\mathcal{Q}}[B]\exp\left[\frac{2\pi ip'}{N}(kB)^TI( kB)\right]e^{i\Omega(X)}~.
\fe
Combining this relation with \eqref{eq:spin_manifold_partition} and \eqref{eq:spin_manifold_equality}, we get the same constraint as in \eqref{eq:LtildeL} so  $\ell$, $\hat p\equiv p/\text{gcd}(p,k)$, $\hat k\equiv k/\text{gcd}(p,k)$ are all co-prime to each other and we further derived
\ie
\tilde p(\tilde p+p')+1=0\text{ mod }\ell~.
\fe
where $\tilde p=\hat k(\hat p)^{-1}_{\hat k\ell}$. Again, because the Gauss sum for the intersection forms of a simply connected smooth spin 4-manifold $X$ is unity, 
 we learned that $e^{i\Omega(X)}=1$.
This concludes the proof.

\section{Spontaneous Symmetry Breaking}\label{app:SSB}

While the main focus of our analysis is on theories with a unique vacuum state on $S^{3}$, it is also possible to spontaneously break duality symmetries leading to multiple vacuum states.  In this case there is no constraint on $N$.  Here we briefly summarize some simple examples of this phenomenon.  We also note that spontaneous duality symmetry breaking is realized in $\mathbb{Z}_{N}$ lattice gauge theory for small $N$ as discussed in section \ref{sec:lattice}.

We can construct an elementary gapped example of spontaneously breaking the duality defect $\mathcal{D}$ by starting from $\mathbb{Z}_{N}$ topological gauge theory (i.e. Dijkgraaf-Witten theory).  This theory has a $\mathbb{Z}_{N}^{(1)}$ one-form symmetry under which the Wilson lines are charged.  In the presence of a background gauge field $B$ the partition function on a simply connected four-manifold $X$ is: 
\begin{equation}\label{pure_theory}
    Z_{\text{DW}}[B] =\frac{1}{\sqrt{|H^{2}(X,\mathbb{Z}_{N})|}} \sum_{c\in H^{2}(X,\mathbb{Z}_{N})}\exp{\bigg[\frac{2 \pi i}{N}\int_{X} c \cup B\bigg]}=\sqrt{|H^{2}(X,\mathbb{Z}_{N})|}\delta(B\text{ mod } N)~.
\end{equation}
Performing the $S$ gauging operation replaces the partition function above by
\begin{equation}
    SZ_{\text{DW}}[B]=\frac{1}{|H^{2}(X,\mathbb{Z}_{N})|} \sum_{b,c\in H^{2}(X,\mathbb{Z}_{N})}\exp{\bigg[\frac{2 \pi i}{N}\int_{X} b\cup (c+B)\bigg]}~.
\end{equation}
The equation of motion for $b$ forces $c = -B$, and as a consequence the theory obtained after gauging is trivial
\begin{equation}
    SZ_{\text{DW}}[B]=Z_{\text{trivial}}[B] = 1~.
\end{equation}
This gauging procedure is reversible: gauging a $\mathbb{Z}_{N}^{(1)}$ one-form symmetry in the trivial theory takes us back to the original Dijkgraaf-Witten theory. 

We can now use these results to construct a theory which is self-dual under gauging $\mathbb{Z}_{N}^{(1)}$ one-form symmetry for all values of $N$. We consider a theory $\mathcal{Q}$ which is the direct sum of the Dijkgraaf-Witten theory and the trivial theory:\footnote{The direct sum of $d$-dimensional theories assigns to a $d-1$ manifold a Hilbert space which is the direct sum of the Hilbert space assigned to each summand, and assigns to a closed $d$ manifold the sum of the partition functions.  See references \cite{Durhuus1994, Sawin1995} for additional details.}
\begin{equation}
    Z_{\mathcal{Q}}\equiv Z_{\text{DW}} \oplus Z_{\text{trivial}}~.
\end{equation}
By construction, $\mathcal{Q}$ has two local ground states 
and correspondingly two topological local operators corresponding to the identity operator in each of the sectors above.  According to the discussion above the $S$ operation exchanges these two summands.  This implies that the theory $\mathcal{Q}$ has a duality defect $\mathcal{D}$ which permutes the two topological local operators.

Physically, the construction above can occur  
at a first order phase transition with a spontaneously broken duality symmetry.  In this case the symmetry defect $\mathcal{D}$ describes a domain wall connecting the two local vacua which are hence related by the gauging operation $S$.  In particular, this is exactly the phase transition which occurs in self-dual $\mathbb{Z}_{N}$ lattice gauge theory for $N \leq 4$. (See section \ref{sec:lattice}.)

The construction above may be straightforwardly generalized to produce myriad examples of spontaneously broken duality symmetry.  Indeed, given any theory $\mathcal{G}$ with $\mathbb{Z}_{N}^{(1)}$ one-form symmetry, we can consider the direct sum:
\begin{equation}
    Z_{\mathcal{Q}}\equiv Z_{\mathcal{G}}\oplus Z_{S\mathcal{G}}~,
\end{equation}
which again realizes spontaneously broken duality symmetry. Similarly, we can construct gapped or gapless phases realizing spontaneous symmetry breaking for the more general non-invertible symmetries arising from invariance under $T^{-p'}S$.  For instance, a spontaneously broken triality symmetry leads in general to three local vacua connected by sequential gauging and stacking operations.  This occurs in the Cardy-Rabinovici model at $\tau_*=e^{\pi i/3}$ for small $N$. (See section \ref{sec:lattice}.)

\let\oldaddcontentsline\addcontentsline% Store \addcontentsline
\renewcommand{\addcontentsline}[3]{}% Make \addcontentsline a no-op
\bibliography{references}
\let\addcontentsline\oldaddcontentsline% Restore \addcontentsline

\end{document}